\documentclass[]{article}

\usepackage[cmex10]{amsmath}
\usepackage{amsfonts}
\usepackage{amssymb}
\usepackage{algorithmic}
\usepackage[caption=false,font=footnotesize]{subfig}
\usepackage{url}
\usepackage[utf8]{inputenc}
\usepackage{graphicx}
\usepackage{cite}
\usepackage{authblk}

\usepackage{siunitx}

\graphicspath{{./figures/}}

\title{Robust Bayesian Filtering and Smoothing Using~Student's $t$ Distribution}

\author[1]{M.~Roth}
\author[1,2]{T.~Ardeshiri}
\author[3]{E.~Özkan}
\author[1]{F.~Gustafsson}
\affil[1]{Dept. of Electrical Engineering, Linköping University, \mbox{SE–581 83 Linköping, Sweden}}
\affil[2]{Dept. of Engineering, Cambridge University, Cambridge CB2 1PZ, UK}
\affil[3]{Dept. of Electrical and Electronics Engineering, Middle East Technical University, Ankara, 06531 Turkey}
\date{\vspace{-1em}\small{michael.roth@liu.se}}

\newcommand{\kpk}{_{k+1|k}}

\newcommand{\kkm}{_{k|k-1}}
\newcommand{\km}{_{k-1}}
\renewcommand{\k}{_k}
\newcommand{\kk}{_{k|k}}
\newcommand{\kL}{_{k|L}}
\newcommand{\kpL}{_{k+1|L}}
\newcommand{\kp}{_{k+1}}
\newcommand{\otk}{_{1:k}}

\newcommand{\otkm}{_{1:k-1}}

\newcommand{\otL}{_{1:L}}

\newcommand{\ztL}{_{0:L}}

\newcommand{\sui}{^{(i)}}

\newcommand{\asf}{\mathsf{a}}

\newcommand{\esf}{\mathsf{e}}

\newcommand{\psf}{\mathsf{p}}

\newcommand{\ssf}{\mathsf{s}}
\newcommand{\tsf}{\mathsf{t}}
\newcommand{\Tsf}{\mathsf{T}}
\newcommand{\vsf}{\mathsf{v}}

\newcommand{\N}{\ensuremath{\mathcal{N}}} 
\newcommand{\G}{\ensuremath{\mathcal{G}}} 
\newcommand{\gvn}{\,|\,} 
\newcommand{\xh}{\hat x}

\newcommand{\yh}{\hat y}

\newcommand{\bpm}{\begin{pmatrix}}
\newcommand{\epm}{\end{pmatrix}}
\newcommand{\bbm}{\begin{bmatrix}}
\newcommand{\ebm}{\end{bmatrix}}
\newcommand{\inv}{^{-1}}

\newcommand{\T}{^{\mathrm{T}}}
\newcommand{\di}{\, \mathrm{d}}
\DeclareMathOperator{\blkdiag}{blkdiag}

\DeclareMathOperator{\cov}{cov}
\DeclareMathOperator{\E}{E}
\DeclareMathOperator{\St}{St}
\DeclareMathOperator{\KL}{KL}

\begin{document}

\maketitle

\begin{abstract}
State estimation in heavy-tailed process and measurement noise is an important challenge that must be addressed in, e.g., tracking scenarios with agile targets and outlier-corrupted measurements. The performance of the Kalman filter (KF) can deteriorate in such applications because of the close relation to the Gaussian distribution. Therefore, this paper describes the use of Student's~$t$ distribution to develop robust, scalable, and simple filtering and smoothing algorithms.

After a discussion of Student's~$t$ distribution, exact filtering in linear state-space models with $t$~noise is analyzed. 
Intermediate approximation steps are used to arrive at filtering and smoothing algorithms that closely resemble the KF and the Rauch--Tung--Striebel (RTS) smoother except for a nonlinear measurement-dependent matrix update. The required approximations are discussed and an undesirable behavior of moment matching for $t$ densities is revealed. A favorable approximation based on minimization of the Kullback-Leibler divergence is presented. 
Because of its relation to the KF, some properties and algorithmic extensions are inherited by the $t$ filter. Instructive simulation examples demonstrate the performance and robustness of the novel algorithms. 
\end{abstract}

\pagebreak
\section{Introduction}

The Kalman filter (KF) is the prevalent tool for estimation in linear state-space models. Its optimality properties as best linear filter in the minimum variance sense are well established~\cite{kailath_linear_2000}. Its derivation as optimal Bayesian filter for Gaussian noise~\cite{sarkka_bayesian_2013} makes it easy to understand. However, the strong connection to the Gaussian distribution also entails some challenges because many real world phenomena cannot be described well by the Gaussian distribution. Examples include measurement outliers produced by unreliable sensors; target maneuvers that can be seen as jump in the process noise; and linearization errors in approximated nonlinear models. A pragmatic way to approach such challenges is to assume heavy-tailed process and measurement noise. We therefore investigate Student's~$t$ distribution as heavy-tailed relative of the Gaussian distribution and its use for filtering and smoothing. 


This paper contributes simple state estimation algorithms in the spirit of the~KF. Our approach is to investigate the Bayesian filtering and smoothing recursions~\cite{sarkka_bayesian_2013} for linear systems with Student's~$t$ noise. Using joint $t$~density approximations, results for the $t$ distribution are then employed to obtain convenient time and measurement updates. The resulting filter appears similar to the KF except for a matrix update that nonlinearly depends on the measurement $y\k$ and the intermediate scaling of matrix parameters. The smoother is equivalent to a Rauch-Tung-Striebel backward pass~\cite{sarkka_bayesian_2013, kailath_linear_2000}. Beyond the algorithm development, the paper contributes a discussion of Student's~$t$ distribution and important differences that must be observed when replacing Gaussian with $t$~noise. The approximation steps of the filter are analyzed and optimal parameter choices with respect to the Kullback-Leibler divergence are derived. Furthermore, an analysis of elliptically contoured distributions shows the origin of the ubiquitous KF expressions. 

The $t$ filter of this paper has first been proposed in~\cite{roth_students_2013,roth_kalman_2013}. This paper fills in many details that were left open in~\cite{roth_students_2013}, e.g., how to best perform the intermediate approximation steps and potential problems with moment matching. Furthermore, we complement the filter with a smoothing algorithm. The discussion of elliptically contoured distributions and the analysis of an exact filtering step for Student's~$t$ noise can serve as basis for algorithm development beyond our proposed solutions. It is important that we consider this work as a stepping stone in the development of more advanced algorithms. 
Therefore, also extensions for the application in nonlinear models are discussed.

Related work on filtering in heavy-tailed noise includes~\cite{masreliez_approximate_1975, masreliez_robust_1977} as early work on measurement outliers; 
\cite{chu_estimation_1973,meinhold_robustification_1989, giron_bayesian_1994} as early approaches based on Student's and elliptically contoured distributions, but without the sequential approximations of our work;
\cite{agamennoni_approximate_2012,piche_recursive_2012} and~\cite{nurminen_robust_2015} as variational approaches for heavy-tailed and skewed measurement noise, respectively; 
\cite{mattingley_real-time_2010} as optimization approach for heavy-tailed measurement noise;
\cite{sornette_kalman-levy_2001,gordon_kalman-levy_2003} as  filters based on the heavy-tailed Lévy distribution; 
\cite{maskell_tracking_2004} as particle filter for heavy-tailed process noise in maneuvering target tracking; 
and \cite{huang_robust_2016-1, tronarp_sigma-point_2016} as sigma point variants of the original $t$ filter~\cite{roth_students_2013} for nonlinear systems.
Smoothing or offline estimation references include~\cite{ardeshiri_approximate_2015,huang_robust_2016} as variational approaches for uncertain process and measurement noise covariance matrices; \cite{gandhi_robust_2010} as batch KF based on robust maximum likelihood estimation; and~\cite{huang_robust_2016-1} as relative of the smoother that we present, but with the repeated use of moment matching. 

The outline of the paper is as follows. Student's~$t$ and the family of elliptically contoured distributions are discussed in Sec.~\ref{sec:distributions}. Bayesian filtering and smoothing in the presence of $t$~noise is discussed Sec.~\ref{sec:filteringT}. A Student's~$t$ filter algorithm is introduced in Sec.~\ref{sec:tFilter} and some of its properties are highlighted in Sec.~\ref{sec:algProp}. A Student's~$t$ smoother is introduced in Sec.~\ref{sec:tSmoother}. Simulation examples are provided in Sec.~\ref{sec:examples} and followed by concluding remarks in Sec.~\ref{sec:conclusions}.

\section{Student's~$t$ and Elliptically Contoured Distributions}\label{sec:distributions}

This section presents useful results for Student's~$t$ distribution. A more detailed account can be found in~\cite{roth_multivariate_2013, roth_kalman_2013}. Some of the insights also hold for the wider class of elliptically contoured distributions, which appear less known in the state estimation context and are therefore included here. 

\subsection{Motivation for using Student's~$t$ distribution}

\begin{figure}[htb]
 \centering
 \includegraphics{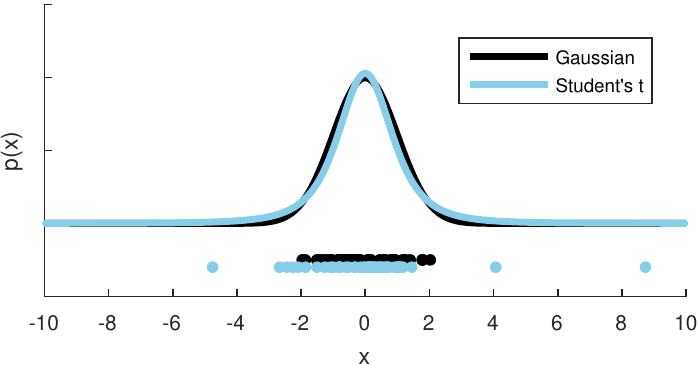}
 \caption{The probability density functions of a $t$ and Gaussian distribution, $\St(0,0.8,3)$ and $\N(0,1)$, respectively. Also shown are $50$ random samples of each.}
 \label{fig:gaussVsT}
\end{figure}

\begin{figure}[htb]
 \centering
 \includegraphics{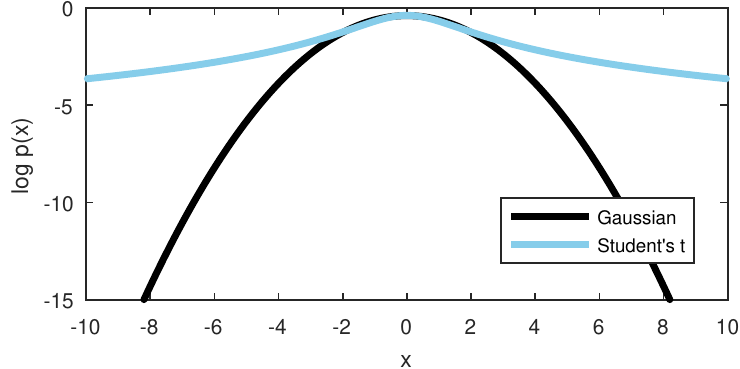}
 \caption{The logarithms of the probability density functions in Fig.~\ref{fig:gaussVsT}.}
 \label{fig:gaussVsTLog}
\end{figure}

The introduction highlighted the need for heavy-tailed distributions to model real-world phenomena. Student's~$t$ distribution is a close relative of the Gaussian distribution that can exhibit heavy-tails for certain parameter choices. Fig.~\ref{fig:gaussVsT} shows the probability density functions $\N(x; 0,1)$ and $\St(x; 0,0.8,3)$ of a Gaussian and a $t$ distribution, respectively. 
Also shown are $50$ realizations of each. The parameters of $\St(0,0.8,3)$ have been chosen to approximately resemble $N(0,1)$ around~$0$. Hence, the samples around~$0$ could come from either of the two distributions. A few $t$ samples, however, are farther from $0$. 
The lack of such values for $\N(0,1)$ lies in the lack of probability mass in the tails. Specifically, the probability of $|x|>3$ is only $0.0027$ for $\N(0,1)$ and $0.044$ for $\St(0,0.8,3)$. The logarithmic illustration in Fig.~\ref{fig:gaussVsTLog} shows how fast $\N(x;0,1)$ decays in comparison to $\St(x;0,0.8,3)$, and visualizes the heavy tails of the latter.

It must be noted that the variances are different ($1$ for $\N(0,1)$ and $3\cdot0.8$ for $\St(0,0.8,3)$). A moment matched $t$~density would be much more peaked around~$0$. This highlights that care must be taken when replacing Gaussian by $t$~noise in estimation problems. 


\subsection{Useful results for Student's~$t$ distribution}

\newcommand{\xDev}{\bbm x_1-\xh_1 \\ x_2-\xh_2 \ebm}
\newcommand{\pOne}{\Sigma_{1}}
\newcommand{\pOneTwo}{\Sigma_{12}}
\newcommand{\pTwo}{\Sigma_{2}}
\newcommand{\pOneGTwo}{\Sigma_{1|2}}

An $n$-dimensional Student's~$t$ random variable $\xi$ is characterized by a mean vector~$\mu$, a positive (semi-)definite symmetric $n\times n$~scale matrix~$\Sigma$, and the scalar degrees of freedom~$\nu>0$. Lower values for~$\nu$ result in heavier tails. For positive definite~$\Sigma$, the probability density function is given by  
%
\begin{equation}\label{eq:tDens}
\St(\xi; \mu, \Sigma, \nu) 
= \tfrac{\Gamma(\frac{\nu + n}{2})}{\Gamma(\frac{\nu}{2})} \tfrac{1}{(\nu\pi)^{\frac{n}{2}}} \tfrac{1}{\sqrt{\det(\Sigma)}} \left(1 + \tfrac{1}{\nu} (\xi-\mu)\T \Sigma\inv(\xi-\mu) \right)^{-\frac{n+\nu}{2}}.
\end{equation}
Alternatively, the density can be written as an infinite Gaussian mixture~\cite{bishop_pattern_2006} with a Gamma distributed latent variable $\lambda$
\begin{equation}\label{eq:tDensMix}
\St(\xi; \mu, \Sigma, \nu) = \int \N(\xi; \mu, \tfrac{1}{\lambda}\Sigma) \G(\lambda; \tfrac{\nu}{2}, \tfrac{\nu}{2}) \di \lambda.
\end{equation}
For $\nu\rightarrow\infty$ the Gamma density tends to a Dirac pulse at $1$ and $\St(x; \mu, \Sigma, \nu)$ converges to $\N(x; \mu, \Sigma)$. 

The covariance matrix of $\xi$ is finite only for $\nu>2$ and given by
\begin{equation}
\cov(\xi) = \tfrac{\nu}{\nu-2} \Sigma, \quad \nu>2.
\end{equation}
Similar conditions apply for the other moments to exist. For example, the case $\nu=1$ yields a Cauchy distribution which does not have a mean value. 

Linear transformations of $t$ vectors maintain their degrees of freedom. The mean and scale matrix are transformed similar to the parameters in the Gaussian case. In particular, for partitioned vectors $\xi$ with
\begin{equation}\label{eq:tDensJoint}
p(\xi) = p(\xi_1, \xi_2) = \St\left(\bbm \xi_1\\ \xi_2 \ebm; \bbm \mu_1\\ \mu_2 \ebm, \bbm \Sigma_1 & \Sigma_{12}\\ \Sigma_{12}\T & \Sigma_2\ebm, \nu\right),
\end{equation}
%
%
the marginal density of $\xi_2$ is given by
\begin{equation}\label{eq:tDensMarg}
p(\xi_2) = \St(\xi_2; \mu_2, \Sigma_2, \nu).
\end{equation}
The conditional density of $\xi_1$ given $\xi_2$ is also a $t$~density, but with increased degrees of freedom. The parameters in
\begin{equation}\label{eq:tDensCond}
p(\xi_1\gvn \xi_2) = \St\left(\xi_1; \mu_{1|2}, \Sigma_{1|2}', \nu_{1|2} \right)
\end{equation}
are given by 
\begin{subequations}\label{eq:condParams}
\begin{align}
\mu_{1|2} &= \mu_1 + \Sigma_{12}\Sigma_{2}\inv(\xi_2-\mu_2) = \mu_1 + \Upsilon(\xi_2-\mu_2), \label{eq:condMean}\\
\Sigma_{1|2} &= \Sigma_1-\Sigma_{12}\Sigma_2\inv\Sigma_{12}\T = \Sigma_1-\Upsilon \Sigma_2 \Upsilon\T, \label{eq:condMatrix}\\
\Sigma_{1|2}' &= \tfrac{\nu + (\xi_2-\mu_2)\Sigma_2\inv(\xi_2-\mu_2)\T}{\nu + n_2} \Sigma_{1|2}, \label{eq:condMatrixNonlin}\\
\nu_{1|2} &= \nu+n_2,
\end{align}
\end{subequations}
where a ``gain matrix'' $\Upsilon=\Sigma_{12}\Sigma_{2}\inv$ has been introduced. The conditional mean value~\eqref{eq:condMean} is the same as in the Gaussian case. The conditional scale matrix~\eqref{eq:condMatrixNonlin} corresponds to that of the Gaussian~\eqref{eq:condMatrix}, but scaled by a factor that depends nonlinearly on $\xi_2$. 
The alert reader will recognize the relation of~\eqref{eq:condMean} and~\eqref{eq:condMatrix} to the Kalman filter measurement update. 

\subsection{Elliptically contoured distributions}\label{sec:distributionsEll}

Both Student's~$t$ and the Gaussian distribution belong to the family of elliptically contoured distributions~\cite{cambanis_theory_1981,fang_symmetric_1989,anderson_introduction_2003}. Common to them is the recurrence of~\eqref{eq:condMean} and~\eqref{eq:condMatrix} and a number of convenient properties.

Elliptically contoured random variables $\xi$ are characterized by probability density functions 
%
%
\begin{equation}\label{eq:densEll}
p(\xi) = \tfrac{1}{\sqrt{\det(\Sigma)}} g\left( (\xi-\mu)\T \Sigma\inv (\xi-\mu)  \right),
\end{equation}
that solely depend on $\xi$ via a quadratic form. Hence, regions of constant $p(\xi)$ are ellipsoids. The function $g(r^2)\ge 0$ is called density generator and must satisfy
\begin{equation}
\idotsint g( u\T u) \di u_1 \dotsm \di u_n = 1
\end{equation}
for~\eqref{eq:densEll} to be a valid probability density function. In the Gaussian case $g(r^2)=(2\pi)^{-\frac{n}{2}} \exp(-r^2/2)$. The Huber cost function in robust regression~\cite{huber_robust_2009} can be related to an elliptically contoured density with $g(r^2)\propto \exp(-r^2)$ for $|r|<r_0$ and $g(r^2)\propto \exp(-|r|)$ otherwise. 

The mean and covariance of $\xi$ are given by~\cite{anderson_introduction_2003}
\begin{subequations}
\begin{align}
\E(\xi) &= \mu,& &\E(r)<\infty,\\
\cov(\xi) &= \frac{\E(r^2)}{n}\Sigma,& &\E(r^2)<\infty,
\end{align}
\end{subequations}
where $r$ has the probability density function
\begin{equation}
p(r) = \frac{2\pi^{\frac{n}{2}}}{\Gamma(\frac{n}{2})} r^{n-1} g(r^2).
\end{equation}

Familiar results can be derived for partitioned parameters as used in~\eqref{eq:tDensJoint}.
Manipulations of the quadratic form using~\eqref{eq:condMean} and \eqref{eq:condMatrix} yield
%
\begin{multline}
(\xi-\mu)\T \Sigma\inv (\xi-\mu)
\\= (\xi_1-\mu_{1|2})\T \pOneGTwo\inv (\xi_1-\mu_{1|2}) + (\xi_2-\mu_2)\T \pTwo\inv (\xi_2-\mu_2), \label{eq:quadForm}
\end{multline}
as shown in App.~\ref{sec:quadForm}. Inserting a known $\xi_2$ does not change that $\xi_1$ enters~\eqref{eq:quadForm} in a quadratic form. Hence, the conditional density $p(\xi_1\gvn \xi_2)$ remains elliptically contoured with
%
%
\begin{subequations}
\begin{align}
\E(\xi_1\gvn \xi_2) &= \mu_{1|2}, \\
\cov(\xi_1\gvn \xi_2) &\propto \Sigma_{1|2}.
\end{align}
\end{subequations}
That is, the conditional mean~\eqref{eq:condMean} is shared by all elliptically contoured distribution. The conditional covariance is always proportional to~\eqref{eq:condMatrix}. From an estimation perspective, the above can be related to the optimality of the Kalman filter measurement update for different noise distributions.

\section{Recursive Bayesian Filtering and Smoothing}
\label{sec:filteringT}

We consider Bayesian filtering and smoothing for linear state-space models
\begin{subequations}\label{eq:linearModel}
\begin{align}
x\kp &= F x\k + v\k,\\
y\k &= H x\k + e\k,
\end{align}
\end{subequations}
where $x\k$ is the $n$-dimensional state and $y\k$ is the $m$-dimensional measurement at time $k$. The initial state $x_0$ and the process and measurement noise, $v\k$ and $e\k$, respectively, are mutually independent and Student's~$t$ distributed with
\begin{subequations}\label{eq:noises}
\begin{align}
p(x_0) &= \St(x_0; \xh_0, P_0, \eta_0), \label{eq:densX0}\\
p(v\k) &= \St(v\k; 0, Q, \gamma),\label{eq:densProcessNoise}\\
p(e\k) &= \St(e\k; 0, R, \delta).\label{eq:densMeasNoise}
\end{align}
\end{subequations}
For white noise $v\k$ and $e\k$, the above forms a Markov model with the transition density and the likelihood 
\begin{subequations}
\begin{align}
p(x\kp\gvn x\k) &= \St(x\kp; Fx\k, Q, \gamma), \label{eq:transitionT}\\
p(y\k\gvn x\k) &= \St(y\k; Hx\k, R, \delta). \label{eq:likelihoodT}
\end{align}
\end{subequations}
A number of useful conditional independence results follow~\cite{sarkka_bayesian_2013}.

From~\eqref{eq:linearModel} and~\eqref{eq:noises} follows that $x\k$ is a random variable for all $k$. Hence, Bayesian state estimation amounts to the challenge of finding conditional densities $p(x\k\gvn y\otL)$ for $k=1,\dotsc,k$, where $y\otL=\{y_1, \dotsc, y_L\}$ contains the available measurements. Three types of problems can be distinguished: $k>L$ is called prediction, $k=L$ filtering, and $k<L$ smoothing.

\subsection{State estimation via transformation, marginalization, and conditioning}

Before delving into recursive Bayesian filtering and smoothing solutions~\cite{sarkka_bayesian_2013} we note that Bayesian state estimation problems can be formulated as basic operations on probability density functions. For example, the probabilistic description in~\eqref{eq:noises} can be used to formulate a density $p(x_0, v_{0:L-1}, e\otL)$ that characterizes all the involved variables to create $x\ztL$ and $y\otL$ in a probabilistic fashion. Such a joint density can also go beyond the independence assumptions of~\eqref{eq:noises} and, for example, account for relations of $e\k$ and $v\k$ in the case of feedback control. 

Furthermore, a linear relation 
\begin{align}
\bbm x\ztL \\ y\otL \ebm &= T \bbm x_0\\ v_{0:L-1}\\ e\otL \ebm
\end{align}
is easily found for linear models~\eqref{eq:linearModel}. Many distributions, including all elliptically contoured distributions of Sec.~\ref{sec:distributions}, have simple expressions for linearly transformed random variables. Hence, a density $p(x\ztL, y\otL)$ can be obtained by transformation of $p(x_0, v_{0:L-1}, e\otL)$. Even the nonlinear case can allow for finding $p(x\ztL, y\otL)$ using a transformation theorem for probability densities~\cite[Theorem 2.1]{gut_intermediate_2009} under some conditions for the involved functions.

From the joint density, the measurements $y\otL$ can be included via the operation of conditioning
\begin{equation}
p(x\ztL\gvn y\otL) = \frac{p(x\ztL, y\otL)}{p(y\otL)}.
\end{equation}

Finally, the marginal smoothing density is obtained by marginalization
\begin{equation}
p(x\k\gvn y\otL) = \idotsint p(x\ztL\gvn y\otL) \di x\otkm \di x_{k+1:L}.
\end{equation}

If carried out exactly, the order of transformation, marginalization, and conditioning can be interchanged. However, care must taken to keep all relevant probabilistic dependencies before marginalization. In linear Gaussian state-space models, all densities commute under the above operations. Hence, algorithms for the case of correlated noise can be easily devised in the above framework. Of course, some extra work is required to arrive at recursive formulas in the spirit of the Kalman filter.

For approximate nonlinear or non-Gaussian state estimation, the recognition of the above operations is often used to devise intermediate approximations that lead to convenient algorithms. For example, nonlinear Kalman filters~\cite{roth_nonlinear_2016} can be derived from an intermediate Gaussian assumption on $p(x\k, y\k\gvn y\otkm)$ from which the Kalman filter measurement update follows via conditioning on $y\k$. In a similar fashion, the filter of Sec.~\ref{sec:tFilter} introduces intermediate $t$ densities. 

\subsection{Sequential solutions}

Compact recursive expressions for Bayesian state estimation~\cite{sarkka_bayesian_2013} can be derived for Markov models specified by a transition density $p(x\kp\gvn x\k)$ and a likelihood function $p(y\k\gvn x\k)$. 

The one-step-ahead prediction and filtering densities are given by
\begin{subequations}\label{eq:bayesFiltSol}
\begin{align}
p(x\kp\gvn y\otk) &= \int p(x\kp\gvn x\k) p(x\k \gvn y\otk)\di x\k, \label{eq:bayesFiltSolPred}\\
p(x\k \gvn y\otk) &= \frac{p(y\k \gvn x\k) p(x\k \gvn y\otkm)}{p(y\k \gvn y\otkm)}, \label{eq:bayesFiltSolCond}
\end{align}
with a normalization constant
\begin{equation}
p(y\k \gvn y\otkm) = \int p(y\k \gvn x\k) p(x\k \gvn y\otkm) \di x\k.
\end{equation}
\end{subequations}

For $L>k$, a backward recursion for the smoothing density is given by 
%
\begin{subequations}\label{eq:bayesSmoothSol}
\begin{align}
p(x\k \gvn y\otL)
&= \int p(x\k, x\kp \gvn y\otL) \di x\kp \label{eq:bayesSmoothSolJoint}\\
&= \int p(x\k \gvn x\kp, y\otk) p(x\kp \gvn y\otL) \di x\kp \label{eq:bayesSmoothSolFactors}\\
&= p(x\k \gvn y\otk)\int \frac{p(x\kp\gvn x\k) p(x\kp\gvn y\otL)}{p(x\kp\gvn y\otk)} \di x\kp. \label{eq:bayesSmoothSolFancy}
\end{align}
\end{subequations}

\subsection{A filtering step for Student's~$t$ noise}
\label{sec:filteringTexact}

We here investigate an exact filtering step for Student's~$t$ noise using the expressions of the previous section. This does not yield a closed form recursion because of a complicated dependence on latent Gamma variables. However, familiar expressions related to the KF equations in App.~\ref{sec:appKf} are revealed.  The following results are the basis for the filter development in Sec.~\ref{sec:tFilter} and can be used to develop approaches beyond it. 

As starting point we assume a Student's~$t$ filtering density
\begin{align}\label{eq:assumedFiltPosterior}
p(x\k\gvn y\otk) &= \St(x\k; \xh\kk, P\kk, \eta\k).
\end{align}
One-step-ahead prediction~\eqref{eq:bayesFiltSolPred} requires the joint density under the integrals. Using the transition density~\eqref{eq:transitionT} and the expression~\eqref{eq:tDensMix} for $t$ densities, we obtain
%
\begin{align}
\nonumber p(x\k, x\kp |y\otk) &= p(x\kp|x\k) p(x\k\gvn y\otk)\\
\nonumber &= \iint \N(x\kp; Fx\k, \tfrac{1}{\lambda'}Q) \N(x\k; \xh\kk, \tfrac{1}{\lambda}P\kk)\\
\nonumber &\qquad \times \G(\lambda'; \tfrac{\gamma}{2}, \tfrac{\gamma}{2})\G(\lambda; \tfrac{\eta\k}{2}, \tfrac{\eta\k}{2}) \di\lambda \di\lambda'\\
\nonumber &= \iint \N\Biggl(\bbm x\k\\ x\kp\ebm; \bbm \xh\kk\\ F\xh\kk \ebm, \bbm \tfrac{1}{\lambda}P\kk & \tfrac{1}{\lambda}P\kk F\T \\ \tfrac{1}{\lambda}F P\kk & \tfrac{1}{\lambda}F P\kk F\T + \tfrac{1}{\lambda'} Q\ebm\Biggr)\\
\nonumber &\qquad \times \G(\lambda'; \tfrac{\gamma}{2}, \tfrac{\gamma}{2})\G(\lambda; \tfrac{\eta\k}{2}, \tfrac{\eta\k}{2}) \di\lambda \di\lambda'\\
&= \iint p(x\k, x\kp|\lambda, \lambda', y\otk) p(\lambda, \lambda'\gvn y\otk) \di\lambda \di\lambda',
\end{align}
which can be split into two factors under the integral. Only the first factor depends on $x\k$. Hence, marginalization of $x\k$ can be performed for the conditionally Gaussian density $p(x\k, x\kp|\lambda, \lambda', y\otk)$ to yield
%
%
%
%
\begin{multline}
p(x\kp|y\otk)\\
= \iint \N(x\kp; F\xh\kk, \tfrac{1}{\lambda}F P\kk F\T + \tfrac{1}{\lambda'} Q) p(\lambda, \lambda'\gvn y\otk) \di\lambda \di\lambda'.
\end{multline}
After introducing
\begin{align}
\xh\kpk &= F\xh\kk,\\
P\kpk(\lambda, \lambda') &= \tfrac{1}{\lambda}F P\kk F\T + \tfrac{1}{\lambda'} Q,
\end{align}
a $(\lambda, \lambda')$-dependent version of the Kalman filter time update~\eqref{eq:kfTime} in App.~\ref{sec:appKf} becomes apparent.

Using the likelihood~\eqref{eq:likelihoodT} and~\eqref{eq:tDensMix}, we can carry out similar steps for the joint prediction density:
%
\begin{align}
\nonumber p(x\k, y\k \gvn y\otkm) &= p(y\k\gvn x\k) p(x\k\gvn y\otkm)\\
%
\nonumber &= \iiint \N(y\k; Hx\k, \tfrac{1}{\lambda''}R) \N(x\k; \xh\kkm, P\kkm(\lambda, \lambda'))\\
\nonumber &\qquad \times p(\lambda, \lambda'\gvn y\otkm) \G(\lambda'';\tfrac{\delta}{2}, \tfrac{\delta}{2}) \di\lambda \di\lambda' \di\lambda''\\
\nonumber &= \iiint p(x\k, y\k| \lambda, \lambda', \lambda'', y\otkm)\\
&\qquad \times p(\lambda, \lambda', \lambda''\gvn y\otkm) \di\lambda \di\lambda' \di\lambda''
\end{align}
to obtain another conditionally Gaussian factor
%
\begin{align}
&\nonumber p(x\k, y\k | \lambda, \lambda', \lambda'', y\otkm)\\
&\quad\quad =\N\Biggl(\bbm x\k \\ y\k\ebm; \bbm \xh\kkm\\ H\xh\kkm \ebm, \bbm P\kkm(\lambda, \lambda') & P\kkm(\lambda, \lambda') H\T \\ HP\kkm(\lambda, \lambda') & S\k(\lambda,\lambda',\lambda'') \ebm\Biggr)\label{eq:tExactJointStateMeas}
\end{align}
under the integrals. Here, a $(\lambda,\lambda',\lambda'')$-dependent version of the Kalman filter residual covariance~\eqref{eq:kfMeasRes} in App.~\ref{sec:appKf} is introduced:
%
%
\begin{equation}
S\k(\lambda,\lambda',\lambda'') = H P\kkm(\lambda, \lambda') H\T + \tfrac{1}{\lambda''}R.
\end{equation}
Again, the density $p(\lambda, \lambda', \lambda''\gvn y\otkm)$ is a product of Gamma densities that does not depend on $x\k$ or $y\k$. Therefore, we can devise a version of the Kalman gain in~\eqref{eq:kfMeasGain},
%
%
\begin{equation}
K\k(\lambda,\lambda',\lambda'') = P\kkm(\lambda,\lambda')H\T S\k(\lambda,\lambda',\lambda'')\inv, \label{eq:tExactGain}
\end{equation}
and perform conditioning on $y\k$ for~\eqref{eq:tExactJointStateMeas}. The resulting
%
\begin{equation}
p(x\k\gvn y\otk,\lambda,\lambda',\lambda'') = \N(x\k; \xh\kk(\lambda,\lambda',\lambda''), P\kk(\lambda,\lambda',\lambda''))
\end{equation}
is specified by the parameters
%
\begin{align}
\nonumber &\xh\kk(\lambda,\lambda',\lambda'') = \xh\kkm 
+ K\k(\lambda,\lambda',\lambda'')(y\k-H\xh\kkm),\\
&P\kk(\lambda,\lambda',\lambda'') = P\kkm(\lambda,\lambda') - K\k(\lambda,\lambda',\lambda'')S\k(\lambda,\lambda',\lambda'')K\k(\lambda,\lambda',\lambda'')\T.
\end{align}
The relation to the Kalman filter measurement update~\eqref{eq:kfMeas} is apparent. 
The complete filtering density is given by 
%
\begin{equation}
p(x\k\gvn y\otk) = \iiint p(x\k\gvn y\otk,\lambda,\lambda',\lambda'') p(\lambda, \lambda', \lambda''\gvn y\otkm) \di\lambda \di\lambda' \di\lambda''. \label{eq:tExactFiltering}
\end{equation}
Because of the latent variables $(\lambda, \lambda', \lambda'')$ and their complicated dependence, the above is not a $t$~density. This discloses a difficult lack of recursive solutions for Student's~$t$ noise. 

A simplistic approximation of~\eqref{eq:tExactFiltering} is to assert that $\lambda=\lambda'=\lambda''$. Then the gain $K\k$ in~\eqref{eq:tExactGain} no longer depends on $\lambda$ and
\begin{equation}
p(x\k\gvn y\otk) = \int  \N(x\k; \xh\kk, \tfrac{1}{\lambda}P\kk) \G(\lambda\gvn \tfrac{\eta\k}{2}, \tfrac{\eta\k}{2}) \di \lambda \label{eq:simplistic}
\end{equation}
is a $t$~density with the Kalman filter mean and covariance~\eqref{eq:kfMeas}. However,~\eqref{eq:simplistic} does not represent any features of the $t$ distribution in the update equations. In contrast, the algorithm of Sec.~\ref{sec:tFilter} proposes intermediate approximations to exploit convenient results for $t$ random variables. Nevertheless, the expressions of this section can serve as starting point for alternative~$t$ filter approaches beyond our development. 


\subsection{A scalar example}\label{sec:filteringTScalar}

The lack of compact recursive solutions for $p(x\k\gvn y\otk)$ in the presence of $t$~noise is a rather sobering result. A theoretical analysis of smoothing would yield similarly complicated expressions.
We here complement these insights with a look at numerical solutions of~\eqref{eq:bayesFiltSol} and~\eqref{eq:bayesSmoothSol} for a linear model, with the result that the filtering and smoothing densities appear unimodal at most time instances. Multimodal filtering densities can appear, though. 

The considered model is a Student's~$t$ random walk that is observed in Student's~$t$ noise,
%
\begin{subequations}\label{eq:modelT}
\begin{align}
x\kp &= x\k + v\k,\\
y\k &= x\k + e\k,
\end{align}
\end{subequations}
%
with $v\k\sim\St(0,1,3)$, $e\k\sim\St(0,1,3)$, and $x_0\sim\St(0,1,3)$. 

\begin{figure}[htp]
 \centering
 \includegraphics{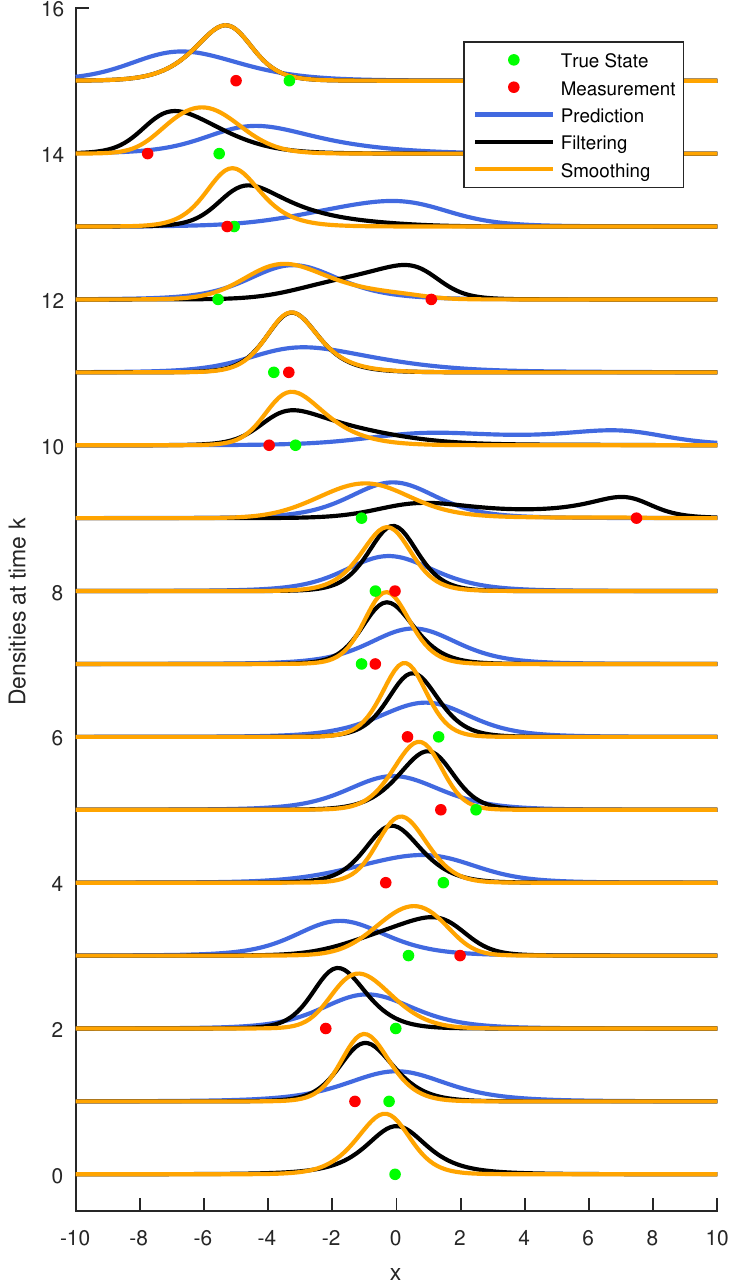}
 \caption{Numerically computed one-step-ahead prediction (blue), filtering (black), and smoothing (orange) densities for a Student's~$t$ random walk observed in $t$~noise for several consecutive time steps. The true states and measurements are illustrated as green and red dots, respectively.}
 \label{fig:studentExact}
\end{figure}

Point mass filters and smoothers~\cite{gustafsson_statistical_2010,roth_computation_2017} are used to compute the Bayesian state estimation densities. The results for a realization with $k=0,\dotsc,15$ are shown in Fig.~\ref{fig:studentExact}.
The densities are unimodal most of the times. The prediction densities $p(x\kp \gvn y\otk)$ appear as broadened versions of $p(x\k\gvn y\otk)$. On two occasions ($k=9$ and $k=12$) the prediction $p(x\k\gvn y\otkm)$ and measurements $y\k$ are in conflict because of an outlier in the latter. The filtering density $p(x\k\gvn y\otk)$ becomes bimodal for $k=9$ and heavily skewed for $k=12$. The smoothing results stem from a backward pass initialized with the filtering result for $k=L=15$. Only one of the two modes persists in $p(x\k\gvn y\otL)$ for $k=9$. 

The numerical results suggest that unimodal, yet heavy-tailed filters and smoothers can provide useful approximations to the exact densities. 

\section{A Student's~$t$ Filter}\label{sec:tFilter}

The filtering algorithm of this section was first presented in~\cite{roth_students_2013}. However, we explain some of the details that were left open in the original submission and highlight potential pitfalls that also affect the recent adaptions in~\cite{huang_robust_2016-1, tronarp_sigma-point_2016}.

\subsection{A simple filter based on intermediate approximations}

Similar to the exact case of Sec.~\ref{sec:filteringTexact}, our starting point is a~$t$~density~\eqref{eq:assumedFiltPosterior} for $p(x\k\gvn y\otk)$. 
The first challenge in the one-step-ahead prediction is that the intermediate joint density $p(x\k,x\kp|y\otk)$ contains a product of $t$ densities that has its origin in the independence of $p(x\k\gvn y\otk)$ and $p(v\k)$ of~\eqref{eq:densProcessNoise}. If, however, we sacrifice the independence and assume a joint $t$~density 
%
\begin{equation}\label{eq:assumedStateProcessNoise}
p(x_k, v\k \gvn y\otk) = \St\left( \bbm x_k\\ v\k \ebm; \bbm \xh\kk\\ 0 \ebm, \bbm P'\kk & 0\\ 0 & Q'\ebm, \eta_k'\right)
\end{equation}
with uncorrelated $x\k$ and $v\k$, joint degrees of freedom $\eta_k'$, and parameters~$P'\kk$ and~$Q'$, then
%
\begin{equation}\label{eq:assumedJointPred}
p(x_k, x\kp \gvn y\otk) = \St\left( \bbm x_k\\ x\kp \ebm; \bbm \xh\kk \\ \xh\kpk \ebm, \bbm P'\kk & P'\kk F\T\\ FP'\kk & P\kpk\ebm, \eta_k'\right)
\end{equation}
follows from the rules for linear transformation of $t$ vectors. The prediction density $p(x\kp \gvn y\otk)=\St(x\kp; \xh\kpk, P\kpk, \eta\k')$, with the parameters
\begin{subequations}
\begin{align}
\xh\kpk &= F\xh\kk,\\
P\kpk &= FP'\kk F\T + Q',
\end{align}
\end{subequations}
follows immediately. The prediction parameters resemble the KF time update~\eqref{eq:kfTime}. However, $P\kpk$ is interpreted as scale rather than covariance matrix here. 

The choice of the adjusted parameters in~\eqref{eq:assumedStateProcessNoise}, marked with primes, is postponed to a later section. However, one choice is $\eta\k'=\min(\eta\k, \gamma)$ to preserve the heaviest tails among the posterior $x\k$ and $v\k$. Choosing $Q'=Q$ and $P'\kk=P\kk$ is then a conservative choice in the sense that the assumed marginal covariance matrices are greater or equal to the original covariance matrices, i.e.,
\begin{equation}
\frac{\eta'_k}{\eta'_k-2}P'\kk\geq \frac{\eta_k}{\eta_k-2}P\kk, \quad \eta_k'\leq \eta_k.
\end{equation}

In order to prepare the measurement update, we must combine a prediction 
\begin{equation}
p(x\k \gvn y\otkm) = \St(x\k; \xh\kkm, P\kkm, \eta\km')
\end{equation}
with $p(e\k)$ of~\eqref{eq:densMeasNoise}. A joint $t$ approximation similar to~\eqref{eq:assumedStateProcessNoise} is given by
%
\begin{equation}\label{eq:assumedStateMeasurementNoise}
p(x_k, e\k \gvn y\otkm) = \St\left( \bbm x\k\\ e\k \ebm; \bbm \xh\kkm\\ 0 \ebm, \bbm P'\kkm & 0\\ 0 & R'\ebm, \eta\k''\right),
\end{equation}
with joint degrees of freedom $\eta\k''$ and adjusted matrices $P'\kkm$ and $R'$. Again, a simple choice is given by $\eta\k''=\min(\eta\k',\delta)$ to preserve the heaviest tails, and $P\kkm' = P\kkm$ and $R'=R$. Consequently, the prediction density of the state and output can be written as
%
\begin{equation}\label{eq:assumedJointMeas}
p(x_k, y_k \gvn y\otkm) = \St\left( \bbm x\k \\ y\k \ebm; \bbm \xh\kkm\\\yh\k \ebm, \bbm P'\kkm & P'\kkm H\T\\ HP'\kkm & S\k\ebm, \eta\k''\right).
\end{equation}
The output prediction and its covariance (similar to~\eqref{eq:kfMeasRes} in the KF) follow from a linear transformation:
\begin{subequations}\label{eq:tFilterYhS}
\begin{align}
\yh\k &= H\xh\kkm, \\
S\k &= HP'\kkm H\T + R'. \label{eq:tMeasS}
\end{align}
\end{subequations}

Finally, a measurement update can be derived from the conditional $t$~density~\eqref{eq:tDensCond} and its parameters~\eqref{eq:condParams}. Similar to the Kalman filter, a gain matrix
\begin{equation}
K\k = P'\kkm H\T S\k\inv
\end{equation}
is used to create the filtering mean and the matrix in
\begin{subequations}\label{eq:tFilterMeas}
\begin{align}
\xh\kk &= \xh\kkm + K\k (y\k - \yh\k),\\
P\kk'' &= P\kkm - K\k S\k K\k\T.
\end{align}
The result $P\kk''$ is, however, further scaled by a factor that nonlinearly depends on $y\k$. Also, the degrees of freedom increase. The update
\begin{align}
P\kk &= \frac{\eta\k'' + (y\k-\yh\k)\T S\k\inv (y\k-\yh\k)}{\eta\k'' + m} P\kk'', \label{eq:tFilterScaling} \\
\eta\k &= \eta\k'' + m \label{eq:fFilterDof}
\end{align}
\end{subequations}
completes the filter recursion by providing the parameters of
\begin{equation}
p(x\k\gvn y\otk) = \St(x\k; \xh\kk, P\kk, \eta\k).
\end{equation}

The degrees of freedom~\eqref{eq:fFilterDof} increase after the measurement update, but are reduced to $\eta'\k$ as first step in the time update. If it were not for this reduction, the filter would soon converge to a KF. In fact, the scaling of~\eqref{eq:tFilterScaling} becomes $1$ for $\eta\k''\rightarrow\infty$ and only the KF measurement update remains. Hence, the KF is one instance of the above filter. This insight is in accordance with the convergence of Student's~$t$ distribution to the Gaussian for infinite degrees of freedom. 

The two approximations~\eqref{eq:assumedStateProcessNoise} and~\eqref{eq:assumedStateMeasurementNoise} lead to $t$ densities in~\eqref{eq:assumedJointPred} and~\eqref{eq:assumedJointMeas} and a filter that resembles the KF except for the nonlinear dependence of $P\kk$ on $y\k$. Similar approximations are common to many state estimation algorithms. For example, nonlinear Kalman filters~\cite{roth_nonlinear_2016, sarkka_bayesian_2013} assume the prediction and filtering densities to be Gaussian to justify the KF measurement update~\eqref{eq:kfMeas}. The interacting multiple model (IMM) filter~\cite{bar-shalom_estimation_2001} reduces Gaussian mixture densities to single components in order to maintain compact expressions. The choice of intermediate Gaussian or $t$ densities could be replaced by other elliptically contoured density with desired properties. A filter could then be derived from the results in Sec.~\ref{sec:distributionsEll}. Again, such a filter would bare some resemblance to the KF.

A point that has not been clarified is the choice of degrees of freedom for~\eqref{eq:assumedStateProcessNoise} and~\eqref{eq:assumedStateMeasurementNoise}. For the user it is particularly simple to assume joint degrees of freedom for $x_0$, $v\k$, and $e\k$ in~\eqref{eq:noises} from the beginning and to maintain these throughout time. The choice $\eta\k'=\eta_0=\gamma=\delta$ results in $\eta\k''=\eta\k'$. The only remaining adjustment is to go from $\eta\k$ back to $\eta\k'$ after the measurement update. A~discussion that also includes the matrix parameters of~\eqref{eq:assumedStateProcessNoise} and~\eqref{eq:assumedStateMeasurementNoise} is given in the following sections. 

\subsection{Approximation by a joint $t$~density}\label{sec:tFilterApproximationJoint}

\begin{figure}[thbp]
 \centering
 \includegraphics{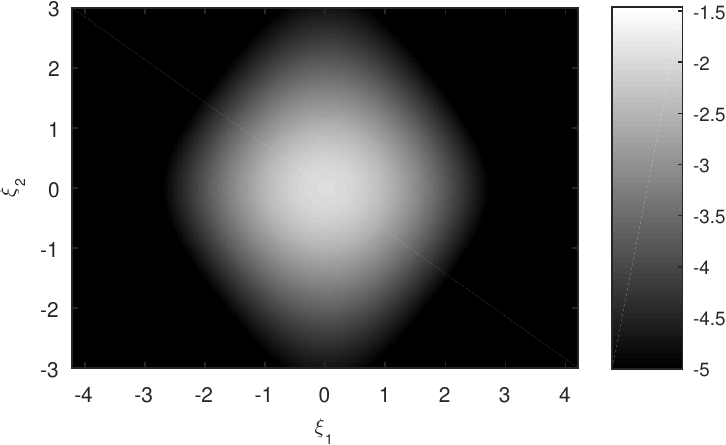}
 \caption{Logarithm of the original density $p(\xi)$ for $\Sigma_1=\Sigma_2=1$, $\nu_1=10$, and $\nu_2=3$. The contours of constant $p(\xi)$ are diamond-shaped.}
 \label{fig:bivariateOrig}
\end{figure}

This section evaluates the approximation of a product of $t$ densities by a joint $t$~density, which is performed in~\eqref{eq:assumedStateProcessNoise} and~\eqref{eq:assumedStateMeasurementNoise} and serves as the basis of the $t$ filter. The discussion is based on a simple experiment in which
\begin{equation}
p(\xi) = p(\xi_1,\xi_2) = \St(\xi_1; 0, \Sigma_1, \nu_1)\St(\xi_2; 0, \Sigma_2, \nu_2)
\end{equation}
with scalar $\xi_1$ and $\xi_2$ is approximated by 
\begin{equation}
q(\xi) = q(\xi_1,\xi_2) = \St\left(\bbm \xi_1\\\xi_2\ebm; \bbm 0\\0\ebm, \bbm \Sigma_1' & 0\\ 0 & \Sigma_2'\ebm, \nu'\right).
\end{equation}
Though not independent any longer, the zero correlation between $\xi_1$ and $\xi_2$ is preserved. 
We chose $\Sigma_1=\Sigma_2=1$, $\nu_1=10$, and $\nu_2=3$. Hence, $\xi_2$ exhibits heavier tails than $\xi_1$. Fig.~\ref{fig:bivariateOrig} shows a plot of $p(\xi)$ on a logarithmic scale. 
The contours of constant $p(\xi)$ are diamond-shaped. For illustration purposes, all values below a threshold of $-5$ are black. 

\begin{figure}[thbp]
 \centering
 \includegraphics{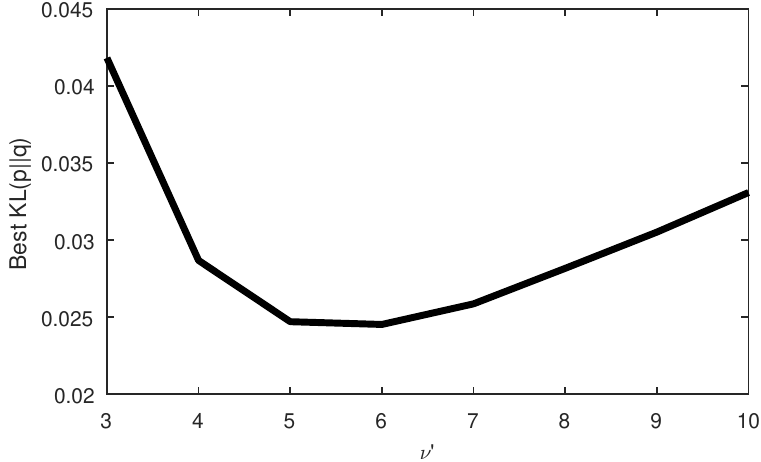}
 \caption{The lowest achieved KLD values for different $\nu'$.}
 \label{fig:bivariateKLD}
\end{figure}

Different parameters $\Sigma_1'$, $\Sigma_2'$, and~$\nu'$ are assessed in terms of the Kullback-Leibler divergence (KLD)
\begin{equation}\label{eq:kld}
\KL(p\Vert q) = \int p(\xi) \log\left(\frac{p(\xi)}{q(\xi)}\right) \di \xi,
\end{equation}
which is a well-established measure for the discrepancy between probability densities~\cite{bishop_pattern_2006}. We compute~\eqref{eq:kld} numerically over a dense grid in $\xi_1$ and $\xi_2$. Furthermore, we restrict our experiments to integer $\nu'\in\{3,\dotsc,10\}$ and change $\Sigma_1'$ and $\Sigma_2'$ in increments of $0.05$ to limit the required computations. The lowest achievable KLD as a function of $\nu'$ is given in Fig.~\ref{fig:bivariateKLD}. The optimal $\nu'=6$ lies between $\nu_1$ and~$\nu_2$. 

\begin{figure}[thb]
 \centering
 \includegraphics{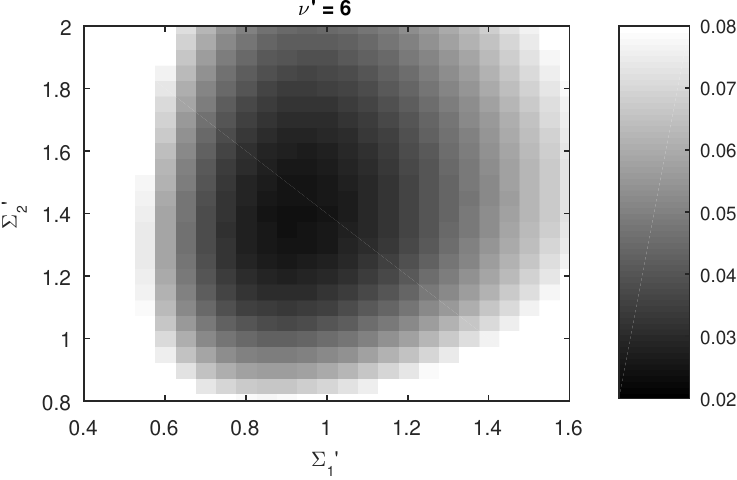}
 \caption{The obtained KLD values for $\nu'=6$ as a function of $\Sigma_1'$ and $\Sigma_2'$.}
 \label{fig:bivariateKldNu6}
\end{figure}

\begin{figure}[thb]
 \centering
 \includegraphics{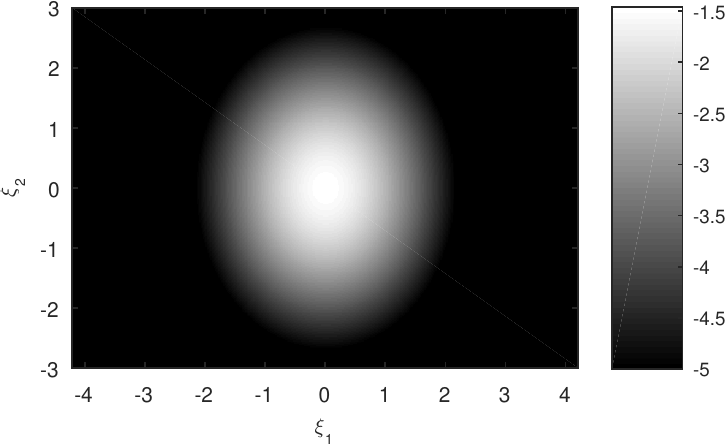}
 \caption{Logarithm of the approximating density $q(\xi)$ for $\Sigma_1'=0.9$, $\Sigma_2'=1.4$, and $\nu'=6$. The contours of constant $q(\xi)$ are ellipses.}
 \label{fig:bivariateBest}
\end{figure}

Fig.~\ref{fig:bivariateKldNu6} illustrates the KLD as a function of $\Sigma_1'$ and $\Sigma_2'$ for $\nu'=6$. 
The best $\Sigma_1'=0.9$ but $\Sigma_1'=\Sigma_1$ is almost as good.
The best $\Sigma_2'=1.4$ is larger than $\Sigma_2$ to account for the increased degrees of freedom. 
The resulting $q(\xi)$ is shown in Fig.~\ref{fig:bivariateBest}.
Although optimal in the KLD sense, some differences to Fig.~\ref{fig:bivariateOrig} should be noted. First, the regions of constant $q(\xi)$ are ellipses. Second, the approximation appears more peaked around~$0$ and the tails in $\xi_2$ are less pronounced. The reason for this is that the KLD~\eqref{eq:kld} is maximized by shifting the probability mass of $q(\xi)$ where $p(\xi)$ is large. 

\begin{figure}[thb]
 \centering
 \includegraphics{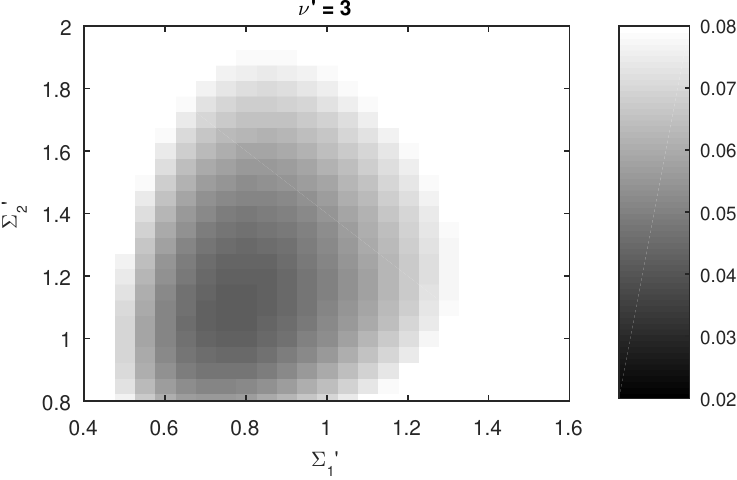}
 \caption{The obtained KLD values for $\nu'=3$ as a function of $\Sigma_1'$ and $\Sigma_2'$.}
 \label{fig:bivariateKldNu3}
\end{figure}
\begin{figure}[thb]
 \centering
 \includegraphics{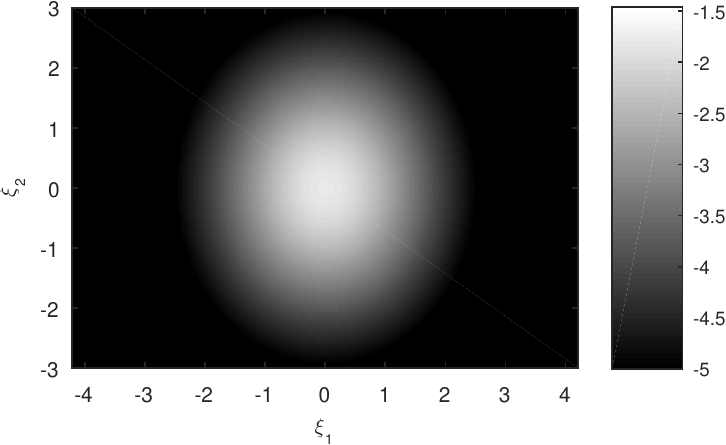}
 \caption{Logarithm of the approximating density $q(\xi)$ for $\Sigma_1'=0.8$, $\Sigma_2'=1.1$, and $\nu'=3$.}
 \label{fig:bivariateAny}
\end{figure}

From a robustness perspective, it is desirable to retain the tails in $\xi_2$. Therefore, we investigate $\nu'=\nu_2=3$ as common degrees of freedom. Although largest in Fig.~\ref{fig:bivariateKLD}, the achievable KLD of $0.04$ for $\nu'=3$ is still on the lower end of the scale in Fig.~\ref{fig:bivariateKldNu6}. The resulting KLD as function of $\Sigma_1'$ and $\Sigma_2'$ is given in Fig.~\ref{fig:bivariateKldNu3}.
The best values are $\Sigma_1'=0.8$ and $\Sigma_2'=1.1$ and the corresponding $q(\xi)$ is shown in Fig.~\ref{fig:bivariateAny}. 
Now, the tails of $p(x)$ are resembled more conservatively. Moreover, the approximation close to the mean appears more similar to $p(x)$ in Fig.~\ref{fig:bivariateOrig}. As a downside, the elliptical approximation makes samples that are large in magnitude for both $\xi_1$ and $\xi_2$ more likely than for the original~$p(x)$. 


To conclude, the decrease in degrees of freedom for $\xi_1$ required a decreased $\Sigma_1'$ to give the optimal KLD. Increasing the degrees of freedom for $\xi_2$ required an increased $\Sigma_2'$. An increase in degrees of freedom for $\xi_2$ leads to less pronounced tails. A conservative choice that preserves these tails, although not optimal in terms of the KLD, would be to set $\nu'=\nu_2$. 

\subsection{Marginal approximation via matrix parameter adjustment}
\label{sec:tFilterApproximationMarginal}

The experiment of Sec.~\ref{sec:tFilterApproximationJoint} investigated a joint $t$~density approximation for a product of two $t$ densities. Although simple for the shown low-dimensional example, this is difficult in general. However, the approximations in Sec.~\ref{sec:tFilterApproximationJoint} were reduced to choosing the parameters of the marginal densities of $\xi_1$ and $\xi_2$ after selecting the common degrees of freedom $\nu'$.

Therefore, we here discuss the approximation of one $n$-dimensional $t$~density $p(\xi)$ by another $t$~density $q(\xi)$ with 
\begin{subequations}
\begin{align}
p(\xi) &= \St(\xi; 0, \Sigma, \nu),\\
q(\xi) &= \St(\xi; 0, c\Sigma, \nu').
\end{align}
\end{subequations}
The scale matrix of $q(\xi)$ is adjusted with a factor $c$ to preserve the correlation among the components of $\xi$. 
Motivated by the intention to keep the heaviest tails, we investigate only reduced degrees of freedom $\nu'<\nu$. 
The following shows that moment matching can entail negative effects and that $c$ can be devised from an offline optimization that does not depend on $\Sigma$. 

Again, our starting point is the Kullback-Leibler divergence~\eqref{eq:kld}. 
We first note that the KLD between $p(\xi)$ and $q(\xi)$ is a function in which~$\xi$ enters only in quadratic forms $\xi\T\Sigma\inv\xi$. Using a technique called stochastic decoupling, which is often applied in nonlinear Kalman filters to simplify Gaussian integrals~\cite{roth_nonlinear_2016}, we obtain
\begin{align}
\nonumber \KL(p\Vert q) 
&= \int \St(\xi; 0, \Sigma, \nu) \log\left( \frac{\St(\xi; 0, \Sigma, \nu)}{\St(\xi; 0, c\Sigma, \nu)} \right) \di \xi\\
&= \int \St(\xi'; 0, I, \nu) \log\left( \frac{\St(\xi'; 0, I, \nu)}{\St(\xi'; 0, cI, \nu)} \right) \di \xi'. \label{eq:kldStandard}
\end{align}
That is, the KLD does does not depend on the specific $\Sigma$, but is a function of $\nu$, $\nu'$, and $n$ only. Furthermore,~\eqref{eq:kldStandard} can be easily evaluated using $n$-dimensional\footnote{Further simplification to obtain a scalar integral is possible~\cite{roth_kalman_2013} but $n$-dimensional Monte Carlo integration works well for our purposes.} Monte Carlo integration for given $c$. 

It follows that $c$ can be found offline by a numerical minimization of~\eqref{eq:kldStandard}. 
\begin{figure}[htbp]
 \centering
 \includegraphics{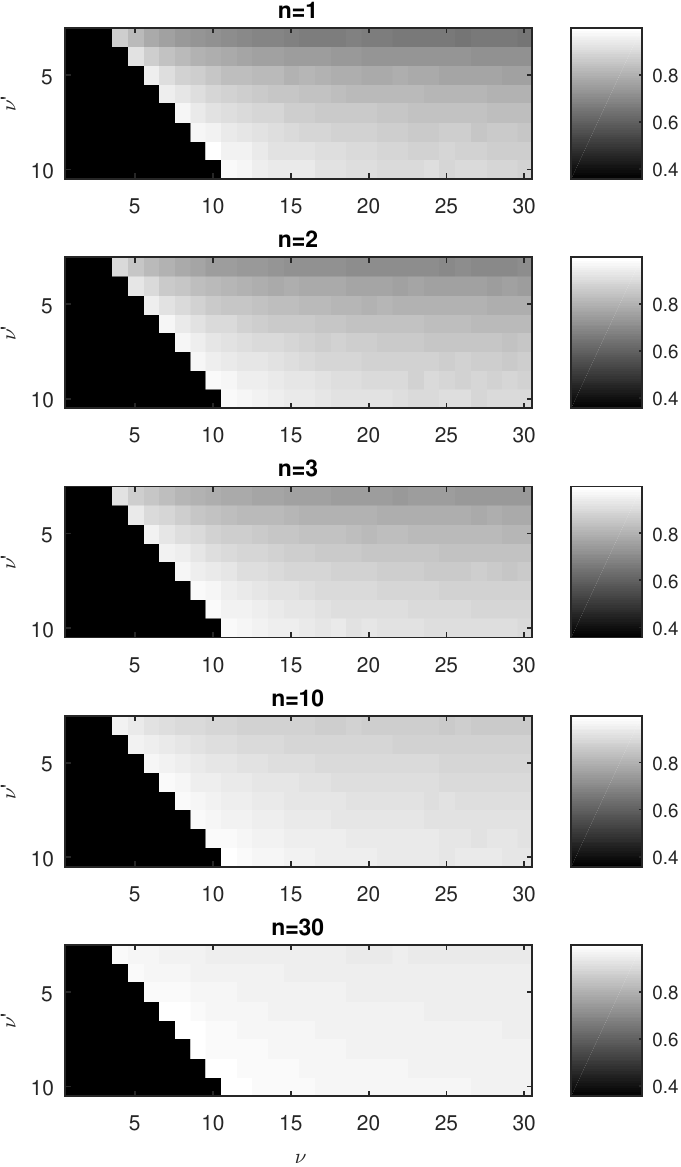}
 \caption{The optimal factors $c$ obtained by KLD minimization for different $n$, $\nu$, and $\nu'$.}
 \label{fig:marginalApproximationKLD}
\end{figure}
Fig.~\ref{fig:marginalApproximationKLD} illustrates the optimal $c$ for different $n$. As $n$ increases, the optimal $c$ approaches $1$ regardless of $\nu$ and $\nu'$. Hence, an adjustment of $\Sigma$ to account for the change from $\nu$ to $\nu'$ is less important in higher dimensions. For $n=1$, we obtain $0.7<c<1$ with smaller $c$ for larger differences $\nu-\nu'$.

\begin{figure}[htbp]
 \centering
 \includegraphics{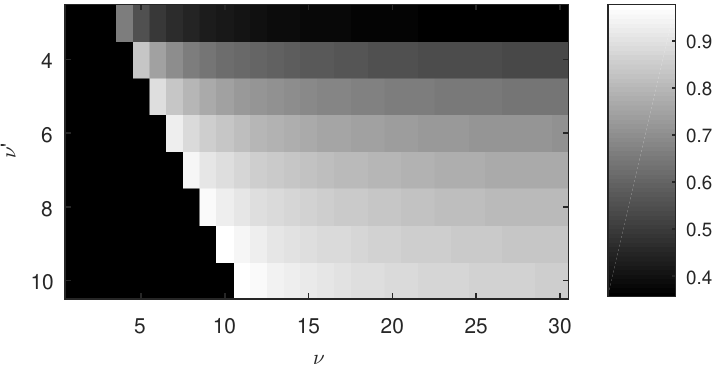}
 \caption{The factors $c$ obtained by moment matching for different $\nu$ and $\nu'$.}
 \label{fig:marginalApproximationMoment}
\end{figure}

As an alternative to the above, we investigate moment matching. In order to preserve the original covariance associated with $p(\xi)$ in $q(\xi)$, we must choose
\begin{equation}\label{eq:cMM}
c = \frac{(\nu'-2)\nu}{\nu'(\nu-2)},
\end{equation}
which can lead to very small values for $\nu'<\nu$. Furthermore, \eqref{eq:cMM} does not depend on $n$. Fig.~\ref{fig:marginalApproximationMoment} illustrates the obtained $c$ for different $\nu$ and $\nu'$. Especially for $\nu'=3$, $c<0.5$ would yield much more peaked density functions. The covariance of a $t$ vector is much influenced by the tails of $p(\xi)$, and the tail behavior depends on~$\nu$. Therefore, moment matching can have the undesirable effect of producing too narrow $q(\xi)$. We illustrate this for a scalar example with $p(\xi)=\N(\xi; 0, 1)$ and $q(\xi)=\St(\xi; 0, c, 3)$, i.e., $\nu=\infty$ and $\nu'=3$. The KLD and moment matching yield $c=0.63$ and $c=1/3$, respectively. The plots of $p(\xi)$ and $q(\xi)$ in Fig.~\ref{fig:marginalApproximationMoment} reveal that the KLD approach better reflects $p(\xi)$. 

\begin{figure}[htb]
 \centering
 \includegraphics{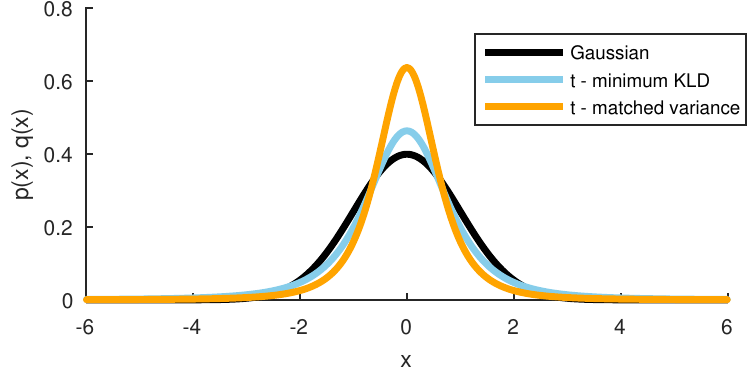}
 \caption{A Gaussian and two $t$~density approximations with $\nu=3$ that are obtained by KLD minimization and moment matching, respectively.}
 \label{fig:gaussTKldMom}
\end{figure}


\section{Algorithmic Properties and Extensions}\label{sec:algProp}

A number of theoretical results as well as ideas for algorithmic extension are provided in this section. 

\subsection{An inherited optimality property}\label{sec:algPropVariance}

We here show that the $t$ filter of Section~\ref{sec:tFilter} can retain the minimum variance optimality of the KF, despite the fundamental difference that $P\kk$ in~\eqref{eq:tFilterScaling} is a nonlinear function of $y\k$. 

For a linear Gaussian model, the Bayesian prediction and filtering densities are Gaussian and their parameters are given by the KF~\cite{sarkka_bayesian_2013}. Now, assume that we employ a $t$ filter step on the correct parameters $\xh\kk$ and $P\kk$ of the KF. For $P\kk'=P\kk$ and $Q'=Q$, the $t$ filter and KF time update yield the same result. After a shift in $k$, we carry out the measurement update. For $P\kkm'=P\kkm$, $R'=R$, and $\eta\k''=\eta\k$, the $t$ filter and KF measurement update differ only by an extra factor
\begin{equation}\label{eq:dFactor}
d(y\k) = \frac{(y\k-\yh\k)\T S\k\inv (y\k-\yh\k) + \eta\k'}{m + \eta\k'}
\end{equation}
in~\eqref{eq:tFilterScaling}. The factor is random because of $y\k$. In the Gaussian case, the exact conditional density of $y\k$ is
\begin{equation}
p(y\k\gvn y\otkm) = \N(y\k; \yh\k, S\k),
\end{equation}
with parameters given by~\eqref{eq:tFilterYhS}.
Consequently, the quadratic form in $d(y\k)$ admits a chi-squared distribution with $m$ degrees of freedom and mean value $m$~\cite{bishop_pattern_2006}. The expected value of~\eqref{eq:dFactor} averaged over all $y\k$ is
\begin{equation}
\E(d(y)) = \frac{m + \eta\k}{m + \eta\k} = 1.
\end{equation}
Hence, the $t$~filter measurement update is that of the KF ``on average'' when applied to a linear Gaussian model. That is, also the minimum variance property of the KF~\cite{gustafsson_statistical_2010} is preserved.

\subsection{Square root implementation}

The similar structure of the $t$ filter in Sec.~\ref{sec:tFilter} and the KF of App.~\ref{sec:appKf} facilitates that also algorithmic reformulations of the latter are inherited. For instance, square root implementations of the $t$ filter can be devised by adjusting square root KF~\cite{gustafsson_statistical_2010, kailath_linear_2000} to include the nonlinear factor in~\eqref{eq:tFilterScaling}. Such a square root $t$ filter then enjoys similar numerical stability and decreases the risk of indefinite matrix parameters. 

\subsection{Application to nonlinear models}

Today's literature offers a wide range of nonlinear KF variants~\cite{roth_nonlinear_2016,sarkka_bayesian_2013} which employ the measurement update~\eqref{eq:kfMeas} of App.~\ref{sec:appKf}  for nonlinear state-space models
\begin{subequations}\label{eq:modelNonlin}
\begin{align}
x\kp &= f(x\k, v\k),\\
y\k &= h(x\k, e\k). 
\end{align}
\end{subequations}
Most of the the nonlinear KF can be derived from an intermediate Gaussian assumption for $p(x\k, y\k \gvn y\otkm)$, with parameters that are computed using linearization~\cite{kailath_linear_2000,gustafsson_statistical_2010,bar-shalom_estimation_2001}, the unscented transformation~\cite{julier_unscented_2004}, numerical integration~\cite{ito_gaussian_2000, arasaratnam_cubature_2009}, or interpolation approaches~\cite{ito_gaussian_2000, norgaard_new_2000}. 

Such approaches can be adapted to the $t$ filter framework by assuming an intermediate $t$~density~\eqref{eq:assumedJointMeas}. A linearization approach similar to the extended KF was suggested in~\cite{roth_students_2013}. First attempts to employ numerical integration and deterministic sampling to find the parameters of~\eqref{eq:assumedJointMeas} can be found in~\cite{huang_robust_2016-1, tronarp_sigma-point_2016}.

We here complement the latter with a Monte Carlo integration scheme. A~time update for~\eqref{eq:modelNonlin} can be achieved by sampling $N$ random state and noise realizations with
\begin{equation}
x\k\sui \sim \St(\xh\kk, P\kk, \eta\k), \quad v\k\sui\sim \St(0, Q, \gamma).
\end{equation}
Then, the state transition function is evaluated for all samples:
\begin{equation}
x\kp\sui = f(x\k\sui, v\k\sui).
\end{equation}
The parameters $\xh\kpk$ and $P\kpk$ of a $t$ prediction density $p(x\kp\gvn y\otk)$ with degrees of freedom $\eta\k'$ can be found using maximum likelihood estimation from the samples. An expectation maximization (EM) algorithm for this can be found in~\cite{mclachlan_em_2008}.

A similar sampling and estimation scheme can be employed to find the remaining parameters of $p(x\k, y\k\gvn y\otkm)$. Again, $N$ samples
\begin{equation}
x\k\sui \sim \St(\xh\kkm, P\kkm, \eta\km'), \quad e\k\sui\sim \St(0, R, \delta)
\end{equation}
are generated, transformed with the measurement function
\begin{equation}
y\k\sui = h(x\k\sui, e\k\sui),
\end{equation}
and processed via the EM algorithm. A $t$ filter measurement update~\eqref{eq:tFilterMeas} concludes the iteration. 

Although heavy in terms of computations, such a Monte Carlo $t$ filter can be especially helpful as benchmark method in the development of less computationally intensive nonlinear $t$ filters~\cite{huang_robust_2016-1, tronarp_sigma-point_2016}.

%
%
%
%

\section{A Student's~$t$ Smoother}\label{sec:tSmoother}

We here develop a backward recursion for Student's~$t$ smoothing. A related approach for nonlinear models is discussed in~\cite{huang_robust_2016-1}, with focus on the required moment integrals. However, the repeated use of moment matching in~\cite{huang_robust_2016-1} can entail some risks, as shown in Sec.~\ref{sec:tFilterApproximationMarginal}. 

Our derivation follows that of the RTS smoother in App.~\ref{sec:appRts}. Starting from the density product in~\eqref{eq:bayesSmoothSolFactors}, we want to arrive at a $t$~density approximation for $p(x\k, x\kp \gvn y\otL)$ in~\eqref{eq:bayesSmoothSolJoint}. Then, simple marginalization yields the smoothing density $p(x\k\gvn y\otL)$. 

The first factor in~\eqref{eq:bayesSmoothSolFactors} follows from the joint $t$~density $p(x\k, x\kp\gvn y\otk)$ in~\eqref{eq:assumedJointPred}. Using the conditioning results for $t$ densities~\eqref{eq:tDensCond} we obtain
\begin{equation}\label{eq:tSmootherCond}
p(x\k\gvn x\kp, y\otk) = \St(x\k; \check x\k, \check P\k, \check \eta\k)
\end{equation}
with 
\begin{subequations}
\begin{align}
\check x\k &= \xh\kk + G\k (x\kp-\xh\kpk),\\
\check P\k' &=  P\kk' - G\k P\kpk G\k\T,\\
\check P\k &= \tfrac{\eta_k' + (x\kp-\xh\kpk)\T P\kpk\inv (x\kp-\xh\kpk)}{\eta_k' + n} \check P\k',\\
\check \eta\k &= \eta\k' + n.
\end{align}
\end{subequations}
and a smoothing gain matrix
\begin{equation}
G\k = P\kk' F\T P\kpk\inv.
\end{equation}
The parameters $P\kk'$, $P\kpk$, and $\eta\k'$ are provided by a previously run $t$ filter.


For the second factor of~\eqref{eq:bayesSmoothSolFactors}, we assume a smoothing density 
\begin{equation}\label{eq:tSmootherDens}
p(x\kp\gvn y\otL) = \St(x\kp; \xh\kpL, P\kpL,\eta\k').
\end{equation}
The degrees of freedom $\eta_k'$ is chosen such that~\eqref{eq:tSmootherCond} and~\eqref{eq:tSmootherDens} could come from conditioning and marginalization of a joint $t$~density $p(x\k, x\kp\gvn y\otL)$, respectively. The $x\k$-parameters of $p(x\k, x\kp\gvn y\otL)$ are the desired smoothing result. 

Similar to the RTS smoother derivation in App.~\ref{sec:appRts}, we revisit the joint, marginal, and conditional $t$ densities in~(\ref{eq:tDensJoint}--\ref{eq:tDensCond}) and identify $\xi_1=x\k$, $\xi_2=x\kp$, $\mu_2=\xh\kpL$, $\Sigma_2=P\kpL$, $\Upsilon=G\k$ to yield algebraic equations for the smoothing parameters $\mu_1=\xh\kL$ and $\Sigma_1=P\kL$. 

The smoothing mean follows immediately:
\begin{equation}\label{eq:tSmootherMean}
\xh\kL = \xh\kk + G\k (\xh\kpL - \xh\kpk).
\end{equation}

In order to find the matrix parameter $P\kL$, we compare 
\begin{subequations}\label{eq:tSmootherTerms}
\begin{align}
\tfrac{\eta\k' + (x\kp-\xh\kpL)\T P\kpL\inv (x\kp-\xh\kpL)}{\eta\k' + n} (P\kL - G\k P\kpL G\k\T)&,\\
\tfrac{\eta\k' + (x\kp-\xh\kpk)\T P\kpk\inv (x\kp-\xh\kpk)}{\eta\k' + n} (P\kk' - G\k P\kpk G\k\T)&.
\end{align}
\end{subequations}
Unfortunately, the dependence on $x\kp$ impedes our efforts. This is not surprising, though, since the product of $t$ densities cannot be turned into a joint $t$~density in general. A rewarding ad-hoc approximation is to ignore the scalar factors in~\eqref{eq:tSmootherTerms}. Then, the matrix update
\begin{equation}\label{eq:tSmootherMatrix}
P\kL = P\kk' + G\k (P\kpk - P\kpL) G\k\T
\end{equation}
follows from simple algebraic manipulations. 

The obtained backward recursion~\eqref{eq:tSmootherMean} and~\eqref{eq:tSmootherMatrix} is identical to the RTS solution~\eqref{eq:rts}. The effect of the omitted factors and the inclusion of potential correction terms remain to be investigated.

\section{Simulation Examples}\label{sec:examples}

\subsection{A scalar example}

We revisit the example of Sec.~\ref{sec:filteringTScalar}. The performance of the KF and the RTS smoother (App.~\ref{sec:appKfRts}) is compared to the Student's~$t$ filter and smoother (Sec.~\ref{sec:tFilter} and~\ref{sec:tSmoother}) for the same trajectory. The exact Bayesian one-step-ahead, filtering, and smoothing densities are shown in Fig.~\ref{fig:studentExact}. The green and red dots mark the true state and measurement, respectively.

\begin{figure}[htp]
 \centering
 \includegraphics{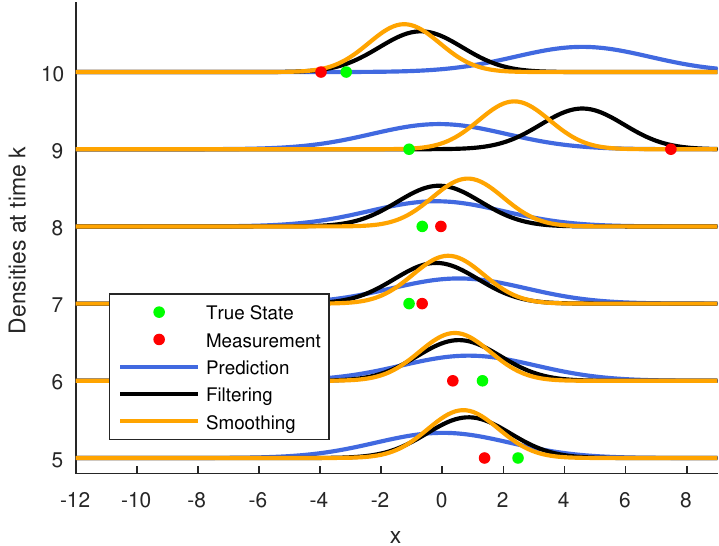}
 \caption{Approximate one-step-ahead prediction (blue), filtering (black), and smoothing (orange) densities obtained from a KF and an RTS smoother on the example of Figure~\ref{fig:studentExact}. The true states and measurements are illustrated as green and red dots, respectively.}
 \label{fig:studentKalman}
\end{figure}

Fig.~\ref{fig:studentKalman} shows the KF and RTS smoother results obtained with the correct variance parameters. Hence, the algorithms are optimal in the minimum variance sense among all linear filters and smoothers. However, an outlier at $k=9$ results in a filtering density that does not cover the true state well. The measurement has a too large influence. The smoothing result for $k=9$ is closer to the true state, but would still yield a worse estimate than the prediction for $k=9$. 

\begin{figure}[htp]
 \centering
 \includegraphics{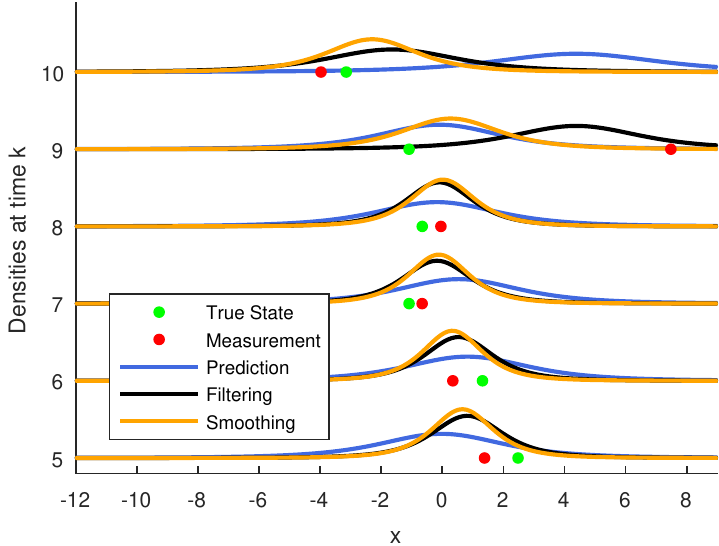}
 \caption{Approximate one-step-ahead prediction (blue), filtering (black), and smoothing (orange) densities obtained from a $t$ filter and smoother on the example of Figure~\ref{fig:studentExact}. The true states and measurements are illustrated as green and red dots, respectively.}
 \label{fig:studentTfilter}
\end{figure}

Fig.~\ref{fig:studentTfilter} shows the performance for the $t$ filter and smoother. No parameter scaling has been employed. Hence, the only difference to the KF is the nonlinear update of $P\kk$ in~\eqref{eq:tFilterScaling}. The outlier at $k=9$ yields a broad filtering density. The smoothing result resembles the prediction again. Hence, the performance of the $t$ algorithms is favorable in comparison to the KF and RTS smoother.

\subsection{Drone tracking}

The $t$ filter and smoother are tested on a hypothetical drone tracking problem. We consider a confined area that is observed by a number of cameras, e.g., a yard of some industry or government building. Fences and walls may keep out intruders on foot, but a drone or unmanned aerial vehicle (UAV) is more difficult to prevent from entering. A tracking filter based on the camera data can be used to, e.g., initiate alarms. 
In comparison to the tracking of larger aircraft, drones are much more agile because of their size and actuation. Position measurements that are obtained from detections by several cameras can be subject to large errors due to the challenges in the image processing, e.g., with moving trees in the background. Thus, the example fits the heavy-tailed noise assumptions of this paper. 

%


Trajectories of maneuvering drones are simulated using a constant velocity model~\cite{gustafsson_statistical_2010} with a four-dimensional state $x\k\T=[\psf\k\T, \vsf\k\T]$ that comprises horizontal position and velocity. The position $\psf\k$ is measured. The state-space model is given by
\begin{subequations}
\begin{align}
x\kp &= \bbm I_2 & \Tsf I_2\\ 0 & I_2 \ebm x\k + \bbm \frac{\Tsf^2}{2}I_2\\ \Tsf I_2 \ebm v\k\\
y\k &= \bbm I_2 & 0 \ebm x\k + e\k.
\end{align}
\end{subequations}
where $\Tsf=0.2$~\si{\second} is the sampling time. The process noise $v\k=\asf\k$ models a zero-mean white acceleration input with a Gaussian distribution $\N(0, Q\k)$. The nominal covariance $Q\k = Q_\mathsf{nom} = I_2/\Tsf^2$ is valid for most $k$. However, maneuvers are introduced by setting $Q\k = Q_\mathsf{man} = 20^2Q_\mathsf{nom}$ for $k=25,75,125$, which corresponds to the times $\tsf=5,15,25$~\si{\second}. The measurements are corrupted by outliers in a similar way. The measurement noise $e\k=\tilde\psf\k$ is zero-mean white Gaussian noise with $\N(0,R\k)$ and covariance $R\k=R_\mathsf{nom}=5^2I_2$ for most $k$ and $R\k=R_\mathsf{out}=25^2$ for $k=50,100$. This induces large measurement errors at the times $\tsf=10,20$~\si{\second}.

\begin{figure}[htp]
 \centering
 \includegraphics{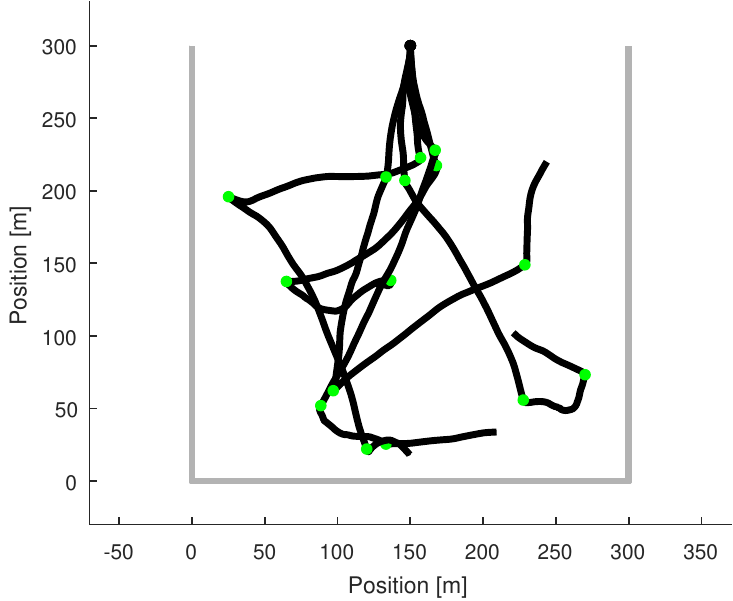}
 \caption{Five simulated drone trajectories. The green dots mark the occurrence of maneuvers.}
 \label{fig:trajectories}
\end{figure}

\begin{figure}[htp]
 \centering
 \includegraphics{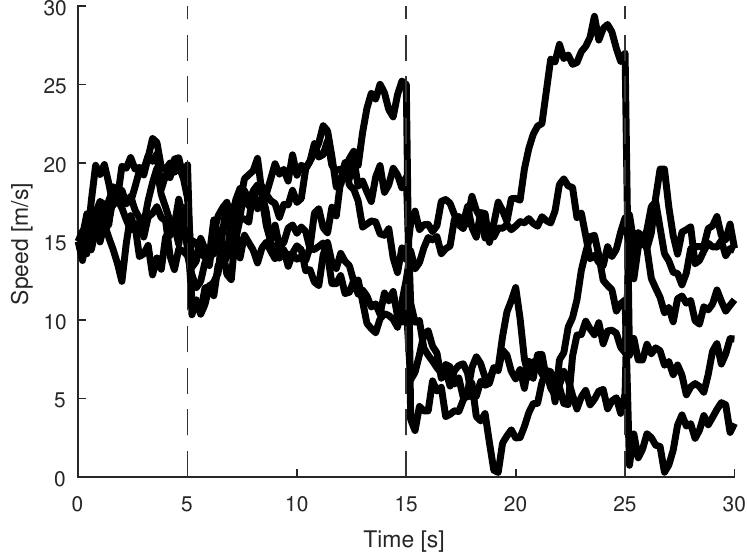}
 \caption{Five simulated speed profiles for the trajectories in Figure~\ref{fig:trajectories}. Maneuver times are illustrated by vertical lines.}
 \label{fig:speed}
\end{figure}

A yard of $300\times 300$~\si{\metre\squared} is considered. Furthermore, the speed $\ssf\k=\| \vsf\k \|_2$ is limited to a maximum of $30$~\si{\metre\per\second}. Trajectories of $151$ time steps ($30$ \si{second}) are simulated with the initial state $x_0=[150, 300, 0, -15]\T$ and only accepted if the position and speed constraints are met at all $k$. Five drone trajectory realizations are illustrated in Fig.~\ref{fig:trajectories}. The green dots mark the occurrence of maneuvers that lead to sudden turns or changes in velocity. Fig.~\ref{fig:speed} illustrates the correspondent speed profiles.

Three filters are compared on the trajectories. The first is a KF with knowledge of the nominal parameters $Q_\mathsf{nom}$ and $R_\mathsf{nom}$ only. Second is a clairvoyant KF that knows also $Q_\mathsf{man}$ and $R_\mathsf{out}$ and the times at which outliers occur. This is the optimal filter for the above problem, but it uses knowledge that is not available in real scenarios. Third is the $t$ filter of Sec.~\ref{sec:tFilter} that employs the nominal parameters only but, assumes $t$~noise with $3$ degrees of freedom. The intermediate approximation steps and the conversion of the noise distributions from Gaussian to Student's~$t$ are carried out via minimization of the KLD, as described in~\ref{sec:tFilterApproximationMarginal}. Furthermore, we run the RTS and $t$ smoothers corresponding to the three filters. 

\begin{figure}[htp]
 \centering
 \includegraphics{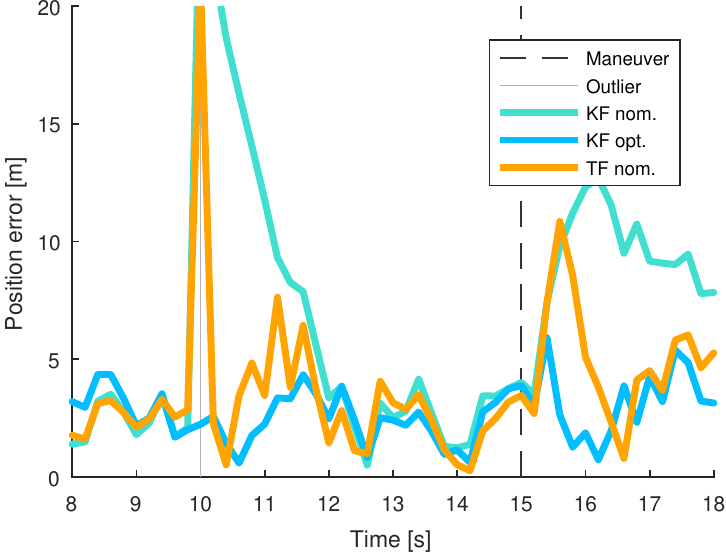}
 \caption{A representative position error profile for the employed filters. Maneuver and outlier times are illustrated by vertical lines.}
 \label{fig:posErr}
\end{figure}

The performance is assessed via the position error $\|\psf\k-\hat\psf\k\|_2$. Fig~\ref{fig:posErr} shows a typical result for an occurring outlier that is followed by a maneuver. The clairvoyant KF is not affected by either. The nominal KF exhibits a large position error that decays slowly. The $t$ filter with nominal parameters also experiences large position errors at the maneuver and outlier times. However, its performance improves quicker than in the nominal~KF. 

\begin{figure}[htp]
 \centering
 \includegraphics{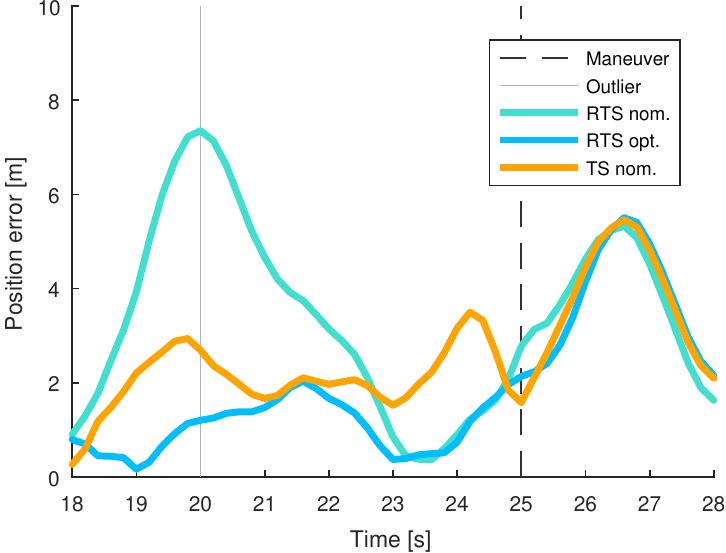}
 \caption{A representative position error profile for the employed smoothers. Maneuver and outlier times are illustrated by vertical lines.}
 \label{fig:posErrSmth}
\end{figure}

Fig.~\ref{fig:posErrSmth} shows a similar result for the smoothers. Again, the clairvoyant RTS performs best. The $t$ smoother is better in the shown example, but performs similar to the nominal RTS in other realizations. The lack of improvement is not surprising since the backward iterations of the RTS and $t$ smoother (see App.~\ref{sec:appRts} and Sec.~\ref{sec:tSmoother}) are algebraically equivalent.  

\begin{figure}[htp]
 \centering
 \includegraphics{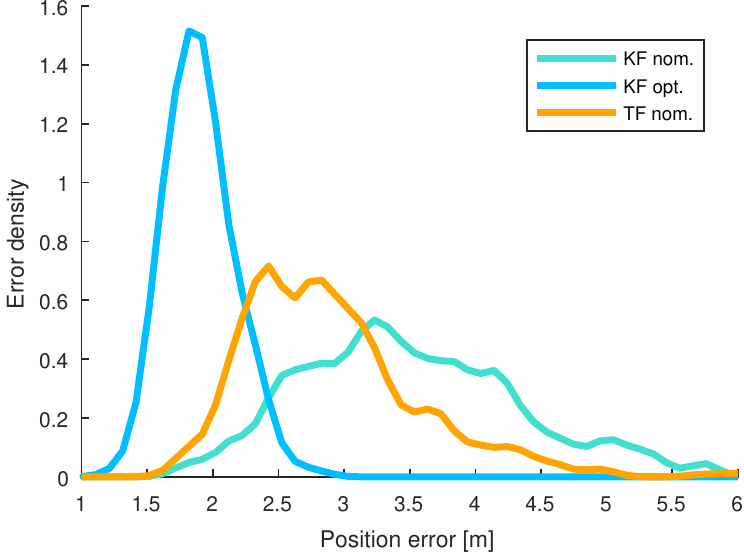}
 \caption{A kernel density estimate of the root mean square position error for the employed filters obtained from $500$ Monte Carlo simulations.}
 \label{fig:mcPosErr}
\end{figure}

To confirm the above results for a larger number of realizations, we perform $500$ Monte Carlo simulations and compute the root mean square error
\begin{equation}
\esf = \left(\tfrac{1}{145} \sum_{k=5}^{150}\|\psf\k-\hat\psf\k\|_2^2\right)^{1/2}
\end{equation}
for each. From the resulting $\esf$, an error density is computed via kernel density estimation\footnote{A kernel density estimate can be interpreted as smoothed histogram.}. The Monte Carlo error for the different filters in shown in Fig.~\ref{fig:mcPosErr}. It can be seen that the $t$ filter improves the result of the nominal KF. 

\begin{figure}[htp]
 \centering
 \includegraphics{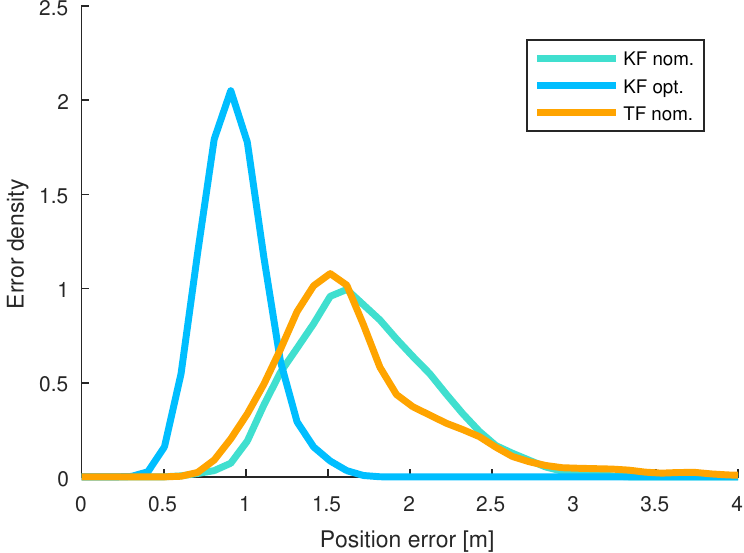}
 \caption{A kernel density estimate of the root mean square position error for the employed smoothers obtained from $500$ Monte Carlo simulations.}
 \label{fig:mcPosErrSmth}
\end{figure}

A similar analysis of the smoothing error in Monte Carlo simulations is illustrated in Fig.~\ref{fig:mcPosErrSmth}. Here, only marginal improvement over the nominal KF can be seen. However, all smoother variants perform well in comparison to the filters, which is a result of the inclusion of more measurements to compute smoothed estimates.

Some further insights from the experiments should be mentioned. An attempt to use gating in the KF, in order to discard measurement outliers, often leads to divergence. The reason for this is that both maneuvers and measurement outliers result in large residuals. Simulations without maneuvers or outliers yield similar performance for all algorithms, which confirms the optimality discussion of Sec.~\ref{sec:algPropVariance}. The choice of degrees of freedom in the $t$ filter appears to have only a~minor effect on the error, unless chosen too large. Also the choice of matrix scaling appears secondary for the $t$ filter. However, the smoother performed worse without the KLD scaling factors. 

The results advocate for the potential of Student's~$t$ filtering and smoothing as simple ways to robustify KF and RTS smoothers. Another simulation study performed by the authors, for a tracking problem without the yard and speed constraints, is described in~\cite[pp.~88--95]{roth_kalman_2013} and confirms the above findings. Furthermore, it compares the $t$ filter to the variational filter of~\cite{agamennoni_approximate_2012} and the particle filter~\cite{sarkka_bayesian_2013, gustafsson_statistical_2010} with the result of a similar advantage of the $t$ filter approach. 

%

\section{Concluding Remarks}\label{sec:conclusions}

We have investigated the use of Student's~$t$ distribution as heavy-tailed alternative to the Gaussian distribition. Results of the Student's~$t$ and other elliptically contoured distributions have been put in context with the ubiquitous expressions of the Gaussian distribution. An exact filtering discussion has shown the lack of exact closed form filtering recursions in linear models with $t$~noise. 

Using two intermediate $t$~density approximations, a simple filter has been derived. The resulting expressions resemble the Kalman filter closely, but include a~nonlinear update of a matrix parameter that depends on the measurement. Furthermore, extensions to nonlinear models and a smoother have been devised. The simulation examples advocate for the potential of our approach as simple but more robust alternative to the Kalman filter and the Rauch-Tung-Striebel smoother. 


\appendix

\section{On Quadratic Forms of Partitioned Vectors}
\label{sec:quadForm}

We here derive the result~\eqref{eq:quadForm}. We assume that $\Sigma$ and $\Sigma_2$ have full rank. The matrix $\Sigma_{1|2}$ of \eqref{eq:condMatrix} is the Schur complement of $\pTwo$ in $\Sigma$. Consequently, $\Sigma\inv$ can be written as~\cite[Equation (A.1.8)]{kailath_linear_2000}
\begin{equation}
\Sigma\inv = \bbm I & 0\\ -\pTwo\inv \pOneTwo\T & I\ebm \bbm \pOneGTwo\inv & 0\\ 0 & \pTwo\inv\ebm  \bbm I & -\pOneTwo\pTwo\inv\\ 0 & I\ebm.
\end{equation}
The matrices $\Sigma\inv$ and $\blkdiag(\pOneGTwo\inv, \pTwo\inv)$ are congruent. Also the determinant $\det(\Sigma)=\det(\pTwo)\det(\pOneGTwo)$ follows. Moreover, 
\begin{align}
\bbm I & -\pOneTwo\pTwo\inv\\ 0 & I\ebm (\xi-\mu) &= \bbm \xi_1 - (\xi_1 + \pOneTwo\pTwo\inv (\xi_2-\mu_2))\\ \xi_2-\mu_2 \ebm
\end{align}
reveals the term $\mu_{1|2}$ of~\eqref{eq:condMean}. The result~\eqref{eq:quadForm} follows after inserting the above into $(\xi-\mu)\T\Sigma\inv (\xi-\mu)$.

\section{The Kalman filter and RTS smoother}\label{sec:appKfRts}

We here list the main equations and refer the reader to the text books~\cite{kailath_linear_2000, gustafsson_statistical_2010, sarkka_bayesian_2013} for further details. 

\subsection{The Kalman filter equations}\label{sec:appKf}

The algorithm is initialized with 
\begin{equation}
\xh_{0|0} = \xh_0, \quad P_{0|0} = P_0.
\end{equation}

The KF time update is given by
\begin{subequations}\label{eq:kfTime}
\begin{align}
\xh\kpk &= F\xh\kk,\\
P\kpk &= FP\kk F\T + Q.
\end{align}
\end{subequations}

The output covariance and Kalman gain
\begin{subequations}
\begin{align}
S\k &= HP\kkm H\T + R, \label{eq:kfMeasRes}\\
K\k &= P\kkm H\T S\k\inv, \label{eq:kfMeasGain}
\end{align}
\end{subequations}
are used to process $y\k$ in the KF measurement update
\begin{subequations}\label{eq:kfMeas}
\begin{align}
\xh\kk &= \xh\kkm + K\k (y\k - H\xh\kkm),\\
P\kk &= P\kkm - K\k S\k K\k\T.
\end{align}
\end{subequations}

\subsection{A compact derivation of the RTS smoother}\label{sec:appRts}

The following derivation is included because of its compactness in comparison to the treatment in, e.g.,~\cite{sarkka_bayesian_2013}, and the analog derivation of the $t$ smoother in Sec.~\ref{sec:tSmoother}.

From the density of a partitioned Gaussian random vector 
\begin{subequations}\label{eq:gaussianResults}
\begin{equation}\label{eq:gaussJoint}
p(\xi_1, \xi_2) = \N\left(\bbm \xi_1\\ \xi_2\ebm; \bbm \mu_1\\ \mu_2\ebm, \bbm \Sigma_1 & \Sigma_{12}\\ \Sigma_{12}\T & \Sigma_2\ebm\right) 
\end{equation}
follow the marginal and conditional densities
\begin{align}
p(\xi_2) &= \N(\xi_2; \mu_2, \Sigma_2),\label{eq:gaussMarg}\\
p(\xi_1\gvn \xi_2) &= \N(\xi_1; \mu_1 + \Upsilon (\xi_2-\mu_2), \Sigma_1 - \Upsilon \Sigma_2\Upsilon\T)\label{eq:gaussCond}
\end{align}
\end{subequations}
with $\Upsilon=\Sigma_{12}\Sigma_{2}\inv$.

In linear Gaussian models~\eqref{eq:linearModel}, the joint prediction density $p(x\k, x\kp\gvn y\otk)$ is also Gaussian with
%
\begin{equation}
p(x\k, x\kp\gvn y\otk) = \N\left(\bbm x\k\\ x\kp\ebm; \bbm \xh\kk\\ \xh\kpk \ebm, \bbm P\kk & P\kk F\T \\ F P\kk & P\kpk \ebm\right),
\end{equation}
where the relation of $P\kpk$ and $P\kk$ is given in~\eqref{eq:kfTime}.
A conditional density
%
\begin{multline}\label{eq:smoothingConditional}
p(x\k\gvn x\kp, y\otk)\\ = 
\N\bigl(x\k;
\xh\kk + G\k (x\kp-\xh\kpk),
P\kk - G\k P\kpk G\k\T\bigr)
\end{multline}
is obtained with the smoothing gain $G\k=P\kk F\T P\kpk\inv$. The product of~\eqref{eq:smoothingConditional} with 
\begin{align}
p(x\kp\gvn y_{1:L}) = \N(x\kp; \xh\kpL, P\kpL)
\end{align}
yields the joint density $p(x\k, x\kp\gvn y_{1:L})$ of~\eqref{eq:bayesSmoothSolJoint}. The product of two Gaussian densities can be shaped into a joint Gaussian density. The procedure is to go from the factors~\eqref{eq:gaussMarg} and~\eqref{eq:gaussCond} to~\eqref{eq:gaussJoint} with the correspondences $\xi_1=x\k$, $\xi_2=x\kp$, $\mu_2=\xh\kpL$, $\Sigma_2=P\kpL$, and $\Upsilon=G\k$, to find $\mu_1=\xh_{k|L}$ and $\Sigma_1=P_{k|L}$. Simple algebraic manipulations yield the backward recursion
\begin{subequations}\label{eq:rts}
\begin{align}
\xh_{k|L} &= \xh\kk + G\k(\xh\kpL - \xh\kpk),\\
P_{k|L} &= P\kk + G\k(P\kpL - P\kpk)G\k\T
\end{align}
\end{subequations}
of the Rauch-Tung-Striebel smoother. The backward pass is initialized with the final KF results $\xh_{L|L}$ and $P_{L|L}$.

\section*{Acknowledgment}

This work was supported by the project Scalable Kalman Filters granted by the Swedish Research Council (VR).

\bibliographystyle{IEEEtran}
\bibliography{student}

\end{document}